\documentclass[prd, superscriptaddress, preprintnumbers, showpacs]{revtex4}
\usepackage{epsfig}
\usepackage{graphicx}
\usepackage{amsmath}
\usepackage{bm}
\usepackage{amssymb}
\usepackage{color}

\newif\ifold             \oldtrue            
\def\ba{\begin{eqnarray}}
\def\ea{\end{eqnarray}}

\newcommand{\be}{\begin{equation}}
\newcommand{\ee}{\end{equation}}
\def\ba{\begin{eqnarray}}
\def\ea{\end{eqnarray}}
\def\bea{\begin{eqnarray}}
\def\eea{\end{eqnarray}}

\begin{document}


\title{Dynamical polarization in ABC-stacked multilayer graphene in a magnetic field}
\date{\today}

\author{O.O. Sobol}
\affiliation{Department of Physics, Taras Shevchenko National Kiev University, 03022, Kiev, Ukraine}

\author{E.V. Gorbar}
\affiliation{Department of Physics, Taras Shevchenko National Kiev University, 03022, Kiev, Ukraine}
\affiliation{Bogolyubov Institute for Theoretical Physics, 03680, Kiev, Ukraine}

\author{V.P. Gusynin}
\affiliation{Bogolyubov Institute for Theoretical Physics, 03680, Kiev, Ukraine}

\begin{abstract}
In the continuum low-energy model, we calculate the one-loop dynamical polarization
functions in  ABC-stacked (rhombohedral) $n$-layer graphene in a magnetic field.
Neglecting the trigonal warping effects, they are derived as functions of wave vector and
frequency at finite chemical potential, temperature, band gap, and the width of Landau levels.
The analytic results are given in terms of digamma functions and generalized Laguerre
polynomials and have the form of double sums over Landau levels. Various particular limits for polarization functions (static, clean, etc.)  are discussed. The intralayer and interlayer screened
Coulomb potentials are numerically calculated as functions of momentum and frequency.
\end{abstract}

\pacs{73.21.Ac, 73.22.Pr}

\maketitle
\section{Introduction}
\label{1}

The physical properties of multilayer graphene strongly depend on the stacking order. ABC-stacked
or rhombohedral graphene is special because, among many possibilities of stacking order, this material
has the flatest electron spectrum at low energy that effectively enhances the role of the electron-electron interactions. According to Refs. \cite{Guinea,Min}, rhombohedral graphene together
with single and bilayer graphene belongs to a new class of two-dimensional electron systems (2DESs)
known as chiral 2DESs \cite{Barlas} due to the chiral properties of its low-energy electron
Hamiltonian. If we neglect the trigonal warping effects (for a discussion of these effects and the electron spectrum in multilayer graphene, see Refs. \cite{Koshino,Sahu}), then the electron energy
in chiral multilayer graphene with $n$ layers at low energy is given by $\varepsilon(\mathbf{p})
\sim |\mathbf{p}|^n$.

While no gap is observed in single-layer graphene at the neutrality point in the absence of
external electromagnetic fields, a gap of $2$ meV is reported in bilayer graphene \cite{Martin,Weitz,Freitag,Velasco}. A much larger gap of room-temperature magnitude
($\sim 42$) meV was recently observed in high-mobility ABC-stacked trilayer graphene \cite{Gillgren}.
A sizable gap can be opened in ABC-stacked trilayer graphene subjected to a perpendicular electric
field \cite{Avetisyan,Gillgren,Bao,Heinz,Sahu,Clapp} (gap opening and gate-tunable band structure
in few-layer graphene are discussed in Refs. \cite{McCann,Russo,Rondeau}).

The screening of the Coulomb potential due to many-body interactions is determined by the
polarization function, which is also an important physical quantity for the spectrum of
collective excitations. The screening  is very essential
in rhombohedral graphene, unlike single-layer graphene, where it plays a relatively minor role.
This is related to the density of states $D(\varepsilon)\sim \varepsilon^{(2-n)/n}$ at energy $\varepsilon$ in gapless graphene, which implies that, as $\varepsilon \to 0$, the density of
states vanishes for single-layer graphene, is constant for bilayer graphene, and diverges for
$n=3$ and higher $n$-layer graphene. These considerations are consistent with the static
polarization function $\Pi(k)$ in undoped gapless ABC-stacked multilayer graphene calculated
in Ref. \cite{Min-polarization}, according to which $\Pi(k) \sim k^{2-n}$ and, consequently,
the static polarization function diverges for $n\ge 3$ as $k \to 0$. The static polarization
function in rhombohedral multilayer graphene was studied also in Ref. \cite{Gelderen}.
The screening effects were taken into account within a simplified model in Ref. \cite{Olsen}.

The dynamical polarization function in ABC-stacked $n$-layer graphene for gapped quasiparticles
in the absence of an external magnetic field was recently numerically calculated in Ref. \cite{Jia}
and its fitting function was found. This polarization function was crucial for the analysis of
the gap generation. Solving the gap equation, it was found that the gap in rhombohedral graphene
attains its maximal value for trilayer graphene and decreases monotonically for $n \ge 4$. Thus,
although the flattening of the low-energy electron bands suggests that the gap should increase
with $n$, the screening effects, which sharply increase with the number of layers $n$, turned
out to be more essential quantitatively.

In the present paper using the effective low-energy model, we study the dynamical polarization
in gapped rhombohedral $n$-layer graphene at finite temperature, chemical potential, impurity rate, quasiparticle gap, and magnetic field.  In the absence of the trigonal warping effects,
we obtain analytical results which, to the best of our knowledge, are not given in the
literature. The paper is organized as follows: We set up the model in Sec. \ref{2}. The dynamical polarization functions in chiral multilayer graphene in a magnetic field are calculated in Sec. \ref{polarization}. The static and dynamical screening in clean rhombohedral graphene is studied
is Sec. \ref{screening}. The main results are summarized and discussed in Sec. \ref{summary}.
Finally, we provide the details of our calculations in Appendixes \ref{3}, \ref{summation}, and \ref{static-lim}. The fermion propagator for low-energy quasiparticles in ABC-stacked multilayer
graphene in a magnetic field is found in Appendix \ref{3}.  The summation over Matsubara frequencies
is performed in Appendix \ref{summation} and some formulas for the polarization functions in the static
limit are given in Appendix \ref{static-lim}.

\section{Model and fermion propagator}
\label{2}

Neglecting the trigonal warping effects, the low-energy electron Hamiltonian in chiral multilayer graphene with $n \ge 2$ layers in a magnetic field is given by \cite{Min,Barlas,Cote,Nakamura}
\be
H_0=\sum_{\xi,s}\int d^2\mathbf{r}\,\Psi^+_{\xi s}(\mathbf{r})
\left( \begin{array}{cc} \Delta_{\xi s} &
- a_n \xi^n  (-i\hat{D}_x-\hat{D}_y)^n \\
- a_n \xi^n  (-i\hat{D}_x+\hat{D}_y)^n  & -\Delta_{\xi s} \end{array} \right)
\Psi_{\xi s}(\mathbf{r})\,,
\label{free-Hamiltonian}
\ee
where $\hat{D}_{i}=\partial_{i}+(ie/c)A_{i}$ ($e>0$) is the covariant derivative with the vector potential $\mathbf{A}=(0,Bx,0)$ which in the Landau gauge describes the magnetic field which points
in the $+z$ direction, $a_n=\gamma_1(\hbar v_F/\gamma_1)^n$,
$v_F \sim 10^6 m/s$ is the Fermi velocity in graphene, and $\gamma_1 \approx 0.39$ eV. Furthermore, $\Delta_{\xi s}$ is the gap generated dynamically or induced by the electric field applied
perpendicular to the planes of graphene. The low-energy effective Hamiltonian (\ref{free-Hamiltonian}) can be utilized for wave vectors $k$ up to $ k_{W}=\gamma_1/(\hbar v_F)$. The two-component spinor field
$\Psi_{\xi s}$ carries the valley ($\xi=\pm$ for the $K$ and $K^{\prime}$ valleys, respectively)
and spin ($s=\pm$) indices. In ABC-stacked multilayer graphene, the low-energy electron states
are located only on the outermost layers, which we will denote as layers $1$ and $n$ in what follows. Furthermore, we use the standard convention for wave functions:
$\Psi_{+s}^T=(\psi_{+A{_1}}, \psi_{+B{_n}})_s$, whereas, $\Psi_{-s}^T = (\psi_{-B{_n}},
\psi_{-A{_1}})_s$. Here, $A_1$ and $B_n$ correspond to those sublattices in the outermost
layers $1$ and $n$, respectively, which are relevant for the low-energy dynamics.
Note that, for $n=2$, Hamiltonian (\ref{free-Hamiltonian}) coincides  with the low-energy Hamiltonian
of bilayer graphene.

The applicability of the two-component model in multilayer graphene is discussed
in  Refs. \cite{Koshino,CoteBarrette}. By computing the eigenstates and Landau-level energies in the tight-binding model and comparing them with the prediction of the two-band model, it was shown \cite{Koshino} that the applicability of the two-band model in trilayer graphene is much restricted due to the rather strong trigonal warping effects which are of the order of $10$\,meV.  Since the experimentally observed gap of $42$\,meV in the absence of external electromagnetic fields in ABC-stacked trilayer graphene is several times larger than the magnitude of the trigonal warping effects, it is natural to expect that the gap will significantly flatten the effects caused by the trigonal warping in trilayer graphene. In our study
we neglect the trigonal warping effects that allows us to get analytical expressions for polarization functions. To account for the trigonal warping requires more numerical computations.

By using $\Psi_{\xi s}(\mathbf{r})=e^{ik y}f_{\xi s}(x)$, the effective Hamiltonian for functions
$f_{\xi s}(x)$ takes the form
\begin{equation}
\tilde{H}_0=\left( \begin{array}{cc} \Delta_{\xi s} &
- \frac{2^{n/2}a_{n} \xi^n}{l^{n}}\hat{b}^{n} \\
- \frac{2^{n/2}a_{n} \xi^n}{l^{n}}(\hat{b}^{\dagger})^{n} & -\Delta_{\xi s}
\end{array} \right),
\label{H-oscillator}
\end{equation}
where the creation and annihilation operators satisfy the standard commutation relation $[\hat{b},\hat{b}^{\dagger}]=1$ and are
\begin{equation}
\hat{b}^{\dagger}=-\frac{i
l}{\sqrt{2}}\left(\partial_{x}-k-x/l^{2}\right),\quad \hat{b}= -\frac{i
l}{\sqrt{2}}\left(\partial_{x}+k+x/l^{2}\right),
\end{equation}
with $l=\sqrt{\hbar c/(eB)}$ being the magnetic length.

We seek the eigenfunctions of Hamiltonian (\ref{H-oscillator}) in the form of the oscillator
wave functions $u_{N}$:
\begin{eqnarray}
f(N,k,x)=\left(\begin{array}{c}c_{1}u_{N-n}(\eta)\\
c_{2}u_{N}(\eta)
\end{array}\right),\quad u_{N}(\eta)=\frac{e^{-\eta^2/2}H_N(\eta)}{\sqrt{2^{N+1}N!\pi^{3/2}l}},
\label{ansatz}
\end{eqnarray}
where $\eta=kl+\frac{x}{l}$, $N \ge n$, $|c_{1}|^{2}+|c_{2}|^{2}=1$, and $H_N(\eta)$ are the
Hermite polynomials. The creation and annihilation operators $b^\dagger$ and $b$ are the
Landau-level ladder operators and they act on the functions $u_{N}(\eta)$ in the standard way.
By using Eqs. (\ref{H-oscillator})-(\ref{ansatz}), we find energy levels at fixed valley and spin:
\begin{equation}
E_{N\alpha}=\alpha M_{N},\quad M_{N}=\sqrt{\Delta^{2}_{\xi s}+\mathcal{E}_{n}^{2} N(N-1)\ldots(N-n+1)},
\,\,\, N \ge n,\,  \alpha=\pm 1,
\end{equation}
where $\mathcal{E}_{n}=\gamma_1\left[{2\hbar^2v^2_F}/{(\gamma^2_1l^2)}\right]^{n/2}$ is the
Landau scale in $n$-layer graphene and eigenfunctions for higher Landau levels (LLs) are given by
\begin{eqnarray}
\Psi_{\xi s}(N,k,\alpha;x,y)=e^{iky}\frac{1}{\sqrt{2M_{N}}}\left(\begin{array}{c}\sqrt{M_{N}
+\alpha\Delta_{\xi s}}\,u_{N-n}(\eta)\\
-(-i)^n \alpha \xi^n \sqrt{M_{N}-\alpha\Delta_{\xi s}}\,u_{N}(\eta)\end{array}\right),\, N\ge n.
\end{eqnarray}
For the lowest Landau level (LLL) $N=0,\ldots,n-1$, we have $E=-\Delta_{\xi s}$ and
the corresponding eigenfunctions are
\begin{equation}
\Psi_{LLL}(N,k,\alpha=-1;x,y)=e^{iky}\left(\begin{array}{c}0 \\
u_{N}(\eta)\end{array}\right),\quad N=0,\ldots,n-1.
\end{equation}
The fermion Green's function satisfies the  equation,
\begin{equation}
(\omega-H_0)S(\mathbf{r},\mathbf{r}^{\prime};\omega)=\delta(\mathbf{r}-\mathbf{r}^{\prime}),
\end{equation}
and can be found through the expansion over the eigenfunctions (see, Appendix \ref{3}).
It has the form
\begin{equation}
S(\mathbf{r},\mathbf{r}^{\prime};\omega)=\exp\left(i\Phi(\mathbf{r},\mathbf{r}^{\prime})\right)
\tilde{S}(\mathbf{r}-\mathbf{r}^{\prime};\omega),
\end{equation}
where $\Phi(\mathbf{r},\mathbf{r}^{\prime})=\frac{ie}{\hbar c}\int
\limits_{\bf r}^{{\bf r}^\prime}A_i^{ext}(z)dz_{i}$ is the Schwinger phase and $\tilde{S}(\mathbf{r}-\mathbf{r}^{\prime};\omega)$ is the
translation-invariant part which is given by Eq. (\ref{translation-invariant-part}).

The electron-electron Coulomb interaction is described by the following interaction
Hamiltonian:
\ba
H_{int}&=&\frac{e^2}{2\kappa}\int\hspace{-1.0mm}
d^3\mathbf{r}\,d^3\mathbf{r}^{\prime}\,\frac{n_{el}(\mathbf{r})n_{el}(\mathbf{r}^{\prime})}
{|\mathbf{r}-\mathbf{r}^{\prime}|},
\label{Coulomb-interaction}
\ea
where $\kappa$ is the dielectric constant,
$n_{el}(\mathbf{r})=\delta(z-(n-1)d/2)\rho_1(\mathbf{x})+\delta(z+(n-1)d/2)\rho_n(\mathbf{x})$
is the three-dimensional electron density in the outermost layers of ABC-stacked $n$-layer
graphene ($d \simeq 0.35$\,nm is the distance between the two neighbor layers), and
two-dimensional charge densities $\rho_1(\mathbf{x})$ and $\rho_n(\mathbf{x})$ in the outermost
layers 1 and $n$ are
\be
\rho_1(\mathbf{x})=\sum_{\xi s}\Psi^+_{\xi s}(\mathbf{x})P_1\Psi_{\xi s}(\mathbf{x})\,,\quad
\rho_n(\mathbf{x})=\sum_{\xi s}\Psi^+_{\xi s}(\mathbf{x})P_n\Psi_{\xi s}(\mathbf{x})\,,
\label{density}
\ee
where $P_1=(1+\xi\tau^3)/2$ and $P_n=(1-\xi\tau^3)/2$ are projectors on states in the layers
$1$ and $n$, respectively. (The appearance of $\xi$ in the projectors is related to our choice
of the form of the spinor $\Psi_{\xi s}$.)

Integrating over $z$ and $z^{\prime}$ in Eq. (\ref{Coulomb-interaction}), one can rewrite
the interaction Hamiltonian as follows:
\ba
H_{int}=\frac{1}{2}\int \hspace{-1.0mm}d^2\mathbf{x}\,d^2\mathbf{x}^{\prime}\left[V(\mathbf{x}-\mathbf{x}^{\prime})
\left(\rho_1(\mathbf{x})\rho_1(\mathbf{x}^{\prime})+\rho_n(\mathbf{x})\rho_n(\mathbf{x}^{\prime})
\right)\hspace{-1.0mm}+2V_{1n}(\mathbf{x}-\mathbf{x}^{\prime})\rho_1(\mathbf{x})
\rho_n(\mathbf{x}^{\prime})
\right].
\label{Coulomb-interaction-1}
\ea
Here, the potential $V(\mathbf{x})$ describes the intralayer interactions and, therefore, coincides
with the bare potential in single-layer graphene whose Fourier transform is given by $V(k)=
{2\pi e^2}/{(\kappa k)}$, $k=|\mathbf{k}|$. The potential $V_{1n}$ describes the interlayer
electron interactions and its Fourier transform is $V_{1n}(k)=
({2\pi e^2}/{\kappa})({e^{-k(n-1)d}}/{k})$. Note that the form of the interaction Hamiltonian (\ref{Coulomb-interaction-1}) coincides with that in bilayer graphene \cite{Jia}.

\section{Dynamical polarization functions and effective interactions}
\label{polarization}

The dynamical polarization determines such physically interesting properties as the
effective electron-electron interaction, the Friedel oscillations, and the spectrum of collective
modes. In multilayer graphene the dynamical polarization functions $\Pi_{jk}$ describe the electron-density correlations on the layers $j,k=1,n$,
\be
\delta(\omega+\omega^{\prime})\delta(\mathbf{k}+\mathbf{k}^{\prime})\Pi_{jk}(\omega,\mathbf{k})=
-i\langle 0|\rho_j(\omega,\mathbf{k})\rho_k(\omega^{\prime},\mathbf{k}^{\prime})|0\rangle \,.
\label{polarization-functions}
\ee
Since the electron states in the low-energy model are located on the outermost layers of ABC-stacked multilayer graphene, there are only two independent functions, $\Pi_{11}=\Pi_{nn}$
and $\Pi_{1n}=\Pi_{n1}$, which describe the correlations of the electron densities on the same
and two different layers. The equalities $\Pi_{11}=\Pi_{nn},\Pi_{1n}=\Pi_{n1}$ are almost evident
from physical equivalence of the outermost layers but can be proved mathematically [see the text
after Eq. (\ref{Pijk}) below].

Taking into account the screening effects, the bare intralayer
$V(k)$ and interlayer $V_{1n}(k)$ electron-electron interactions transform into
\ba
\hat{V}_{\mbox{\scriptsize eff}}=\hat{V}\cdot\frac{1}{1+\hat{V}\cdot\hat{\Pi}}=
\left(\begin{array}{cc}{V}_{\mbox{\scriptsize eff}}&{V}_{1n\,\mbox{\scriptsize eff}}
\\ {V}_{1n\,\mbox{\scriptsize eff}}& {V}_{\mbox{\scriptsize eff}}\end{array}\right),
\quad \hat{V}=\left(\begin{array}{cc}{V}&{V}_{1n}
\\ {V}_{1n}& {V}\end{array}\right),\quad \hat{\Pi}=\left(\begin{array}{cc}
\Pi_{11}&\Pi_{1n}\\ \Pi_{1n}& \Pi_{11}\end{array}\right),
\ea
with
\ba
&&{V}_{\mbox{\scriptsize eff}}(\omega,k)=\frac{2\pi e^2}{\kappa}\,\frac{k+\frac{2\pi
e^2}{\kappa}\Pi_{11}(1-e^{-2(n-1)kd})} { \left[k+\frac{2\pi e^2}{\kappa}(\Pi_{11}+\Pi_{1n})
(1+e^{-(n-1)kd})\right]\left[k+\frac{2\pi e^2}{\kappa}(\Pi_{11}-\Pi_{1n})
(1-e^{-(n-1)kd})\right]}\,, \label{interaction-effective}\\
&& {V}_{1n\,\mbox{\scriptsize eff}}(\omega,k)=\frac{2\pi e^2}{\kappa}\,\frac{ke^{-(n-1)kd}
-\frac{2\pi e^2}{\kappa}\Pi_{1n}(1-e^{-2(n-1)kd})} {\left[k+\frac{2\pi e^2}{\kappa}
(\Pi_{11}+\Pi_{1n})(1+e^{-(n-1)kd})\right]\left[k+\frac{2\pi e^2}{\kappa}(\Pi_{11}-\Pi_{1n})
(1-e^{-(n-1)kd})\right]}\,.
\label{interaction-ND}
\ea
Since $\Pi_{11}$ and $\Pi_{1n}$ depend on $\omega$, the effective interactions
${V}_{\mbox{\scriptsize eff}}$ and ${V}_{1n\,\mbox{\scriptsize eff}}$ depend on it, too.
Note that  the form of the effective interactions (\ref{interaction-effective}) and (\ref{interaction-ND}) in rhombohedral graphene in the low-energy model is the same as in bilayer graphene (see, e.g., Appendix A in Ref. \cite{bilayer}).

Our low-energy model is valid up to the ultraviolet cutoff $k_{W}=\gamma_1/(\hbar v_F)$. Since
$k_{W}d=0.2$, we can expand for $n \le 3$ the exponentials in Eqs. (\ref{interaction-effective})
and (\ref{interaction-ND}) in the Taylor series in $k$ for all $k$ up to the cutoff. Then,
retaining only the zero and first terms in these expansions, we obtain the following effective interactions which are expected to be good approximations in the infrared (IR) region for all
$n$ rather than only for $n \le 3$:
\begin{eqnarray}
&&{V}^{IR}_{\mbox{\scriptsize eff}}(\omega,k)=\frac{2\pi e^2}{\kappa}\,\frac{1+\frac{4\pi
e^2}{\kappa}\Pi_{11}(n-1)d} {\left[k+\frac{4\pi e^2}{\kappa}\Pi
\right]\left[1+\frac{2\pi e^2}{\kappa}(\Pi_{11}-\Pi_{1n})(n-1)d\right]}\,, \label{interaction-effective_simp2}\\
&& {V}^{IR}_{1n\,\mbox{\scriptsize eff}}(\omega,k)=\frac{2\pi e^2}{\kappa}\,\frac{1
-\frac{4\pi e^2}{\kappa}\Pi_{1n}(n-1)d} {\left[k+\frac{4\pi e^2}{\kappa}
\Pi\right]\left[1+\frac{2\pi e^2}{\kappa}(\Pi_{11}-\Pi_{1n})(n-1)d\right]}\,,
\label{interaction-ND_simp2}
\end{eqnarray}
where
\be
\Pi(\omega,\mathbf{k})\equiv \Pi_{11}(\omega,\mathbf{k})+\Pi_{1n}(\omega,\mathbf{k}).
\label{pfdefed}
\ee
It was noted in Ref. \cite{Jia} when studying the gap generation in chiral multilayer graphene
in the absence of a magnetic field that the polarization functions in the numerator and
in the second square brackets in the denominator of the effective interactions (\ref{interaction-effective}) and (\ref{interaction-ND}) somewhat
compensate each other. Therefore, we may try to further simplify the effective interactions (\ref{interaction-effective_simp2}) and
(\ref{interaction-ND_simp2}) by approximating them by
\begin{equation}
\label{interaction-effective_simp3}
{V}^{appr}_{\mbox{\scriptsize eff}}(\omega,k)={V}^{appr}_{1n\,\mbox{\scriptsize eff}}(\omega,k)=\frac{2\pi e^2}{\kappa}\,\frac{1}
{\left[k+\frac{4\pi e^2}{\kappa}\Pi\right]}.
\end{equation}
The approximate expressions (\ref{interaction-effective_simp2}),(\ref{interaction-ND_simp2}),
and (\ref{interaction-effective_simp3}) will be discussed and compared with the effective
interactions (\ref{interaction-effective}) and (\ref{interaction-ND}) in the next section
after we calculate the polarization functions.

The dynamical polarization functions are given by
\begin{equation}
\Pi_{jk}(\omega, \mathbf{r})=i \int\limits_{-\infty}^{\infty}\frac{d \omega^{\prime}}{2\pi}
{\rm Tr}\left[P_{j}\tilde{S}(\mathbf{r},\omega^{\prime})
P_{k}\tilde{S}(-\mathbf{r},\omega^{\prime}-\omega)\right],
\label{Pijk}
\end{equation}
where the trace is taken over spinor, valley, and spin indices.
The scalar functions $\Pi_{jk}(\omega,\mathbf{r}))$ depend on $\mathbf{r}^2$. Since the trace
${\rm Tr}$ in explicit expressions for $\Pi_{jk}$ includes the summation over the valley index
$\xi=\pm$, one can make the change $\xi\to-\xi$, then the projector $P_1$ changes to $P_n$
[see definitions of the projectors $P_1, P_n$ after Eq. (\ref{density})]. Furthermore, the definition of
the propagator Eq. (A5) implies in the case of valley independent gaps that $\tilde{S}(\omega,\mathbf{r},-\xi)= \tilde{S}(\omega,-\mathbf{r},\xi)$, hence we obtain the symmetry properties $\Pi_{11}(\omega,\mathbf{r})=\Pi_{nn}(\omega,\mathbf{r}), \Pi_{1n}(\omega,\mathbf{r})=\Pi_{n1}(\omega,\mathbf{r})$. This symmetry breaks for gaps
depending on the valley index $\xi$ (like $\Delta_{\xi s}=\xi\Delta$ but not $\Delta_{\xi s}=
\xi s\Delta$), and for simplicity, in what follows
we consider (if not stated otherwise) a gap which does not depend on valley and spin indices,
$\Delta_{\xi s}\equiv\Delta$. This is the so-called Haldane gap \cite{Haldane} and
the corresponding generalization to valley and spin-dependent gaps is straightforward.

Taking into account Eq. (\ref{translation-invariant-part}), we obtain ($z=\mathbf{r}^2/2l^2$)
\begin{eqnarray}
\Pi_{11}(\omega, \mathbf{r})=\frac{2i e^{-z}}{(2\pi l^{2})^{2}}\hspace{-1mm} \int\limits_{-\infty}^{\infty}\hspace{-1mm} \frac{d\omega^{\prime}}{2\pi}\hspace{-1mm} \sum\limits_{N,N^{\prime}=0}^{\infty}
\frac{(\omega^{\prime}-\Delta)(\omega^{\prime}-\omega-\Delta)L_{N}(z)L_{N^{\prime}}(z)
+(\omega^{\prime}+\Delta)(\omega^{\prime}-\omega+\Delta)L_{N-n}(z)L_{N^{\prime}-n}(z)}
{(\omega^{\prime 2}-M^{2}_{N})((\omega^{\prime}-\omega)^{2}-M^{2}_{N^{\prime}})},
\label{polarization-coordinate-11}
\end{eqnarray}
\begin{equation}
\label{polarization-coordinate-1n}
\Pi_{1n}(\omega, \mathbf{r})=\frac{2i e^{-z}}{(2\pi l^{2})^{2}} \int\limits_{-\infty}^{\infty}
\frac{d \omega^{\prime}}{2\pi}\sum\limits_{N,N^{\prime}=0}^{\infty}\frac{2a^{2}_{n}}{l^{4n}
(\omega^{\prime 2}-M^{2}_{N})((\omega^{\prime}-\omega)^{2}-
M^{2}_{N^{\prime}})}r^{2n}L^{n}_{N-n}(z)L^{n}_{N^{\prime}-n}(z),
\end{equation}
where the functions $L^\alpha_N (x)$ are generalized Laguerre polynomials;
by definition, $L_N(x)=L^0_N(x)$, $L^\alpha_{N-n} (x)=0$ and $M_N=|\Delta|$ if $N<n$.
Using Eq. (\ref{second-integral}) in Appendix \ref{summation}, we perform the Fourier
transform and find the following dynamical polarization functions in momentum space:
\begin{eqnarray}
\Pi_{11}(\omega, \mathbf{k})&=&\frac{i}{\pi l^{2}} \int\limits_{-\infty}^{\infty}
\frac{d \omega^{\prime}}{2\pi}\sum\limits_{N,N^{\prime}=0}^{\infty}
\frac{(-1)^{N+N^{\prime}}e^{-y}}{(\omega^{\prime 2}-M^{2}_{N})
((\omega^{\prime}-\omega)^{2}-M^{2}_{N^{\prime}})}\nonumber\\
&\times&\left[(\omega^{\prime}-\Delta)(\omega^{\prime}-\omega-\Delta)L^{N-N^{\prime}}_{N^{\prime}}
(y)L^{N^{\prime}-N}_{N}(y)\right.\nonumber\\
&+&\left.(\omega^{\prime}+\Delta)(\omega^{\prime}-\omega+\Delta)L^{N-N^{\prime}}_{N^{\prime}-n}
(y)L^{N^{\prime}-N}_{N-n}(y)\right],\quad y=\frac{\mathbf{k}^2 l^2}{2},
\label{polarization-imp-11}\nonumber\\
\label{polarization-imp-1n}
\Pi_{1n}(\omega, \mathbf{k})&=&\frac{i}{\pi l^{2}} \int\limits_{-\infty}^{\infty}
\frac{d \omega^{\prime}}{2\pi}\sum\limits_{N,N^{\prime}=0}^{\infty}
\frac{(-1)^{N+N^{\prime}}e^{-y}}{(\omega^{\prime 2}-M^{2}_{N})
((\omega^{\prime}-\omega)^{2}-M^{2}_{N^{\prime}})}\nonumber\\
&\times&2\left[\mathcal{E}^{2}_{n}N(N-1)\ldots(N-n+1)\right]L^{N^{\prime}-N}_{N}
(y)L^{N-N^{\prime}}_{N^{\prime}-n}(y).
\end{eqnarray}
At finite temperature and chemical potential the integration over $\omega^{\prime}$ is
replaced by the sum over Matsubara frequencies $\int d\omega^{\prime} \rightarrow
2\pi iT\sum\limits_{m=-\infty}^{+\infty}$, $\omega^{\prime}\,\rightarrow\,i\omega_{m}+\mu$,
where $\omega_{m}=\pi T (2m+1)$ and $\omega\, \rightarrow\,i\Omega_{p}$, $\Omega_{p}=2p\pi T$.
In order to take into account the finite width $\Gamma_{N}$ of the Landau levels or,
equivalently, the scattering rate of quasiparticles, we replace also $i\omega_{m}\rightarrow i\omega_{m}+i\Gamma_{N}{\rm sgn}\omega_{m}$. The width $\Gamma_{N}$ is expressed through
the retarded fermion self energy and, in general, depends on energy, temperature, magnetic
field, and the Landau-level index. In our calculations, we assume that the width is
independent of energy (frequency) but we keep its dependence on the Landau-level index.
The summation over the Matsubara frequencies in the polarization functions is performed in
Appendix \ref{summation}. Our final analytical expressions for the dynamical polarization
functions after analytical continuation take the form of double sums over Landau levels:
\begin{equation}
\Pi_{11}(\omega,y)=-\frac{e^{-y}}{4\pi l^{2}} \sum\limits_{N,N^{\prime}=0}^{\infty}\frac{(-1)^{N+N^{\prime}}}{M_{N}M_{N^{\prime}}}
\sum\limits_{\lambda,\lambda^{\prime}=\pm 1}\mathcal{S}(N,N',\lambda,\lambda',\omega)\nonumber\\
\end{equation}
\begin{equation}
\times\left[(M_{N}-\lambda\Delta)(M_{N^{\prime}}-\lambda^{\prime}\Delta)L^{N-N^{\prime}}_{N^{\prime}}
(y)L^{N^{\prime}-N}_{N}(y)+(M_{N}+\lambda\Delta)(M_{N^{\prime}}+\lambda^{\prime}\Delta)
L^{N-N^{\prime}}_{N^{\prime}-n}(y)L^{N^{\prime}-N}_{N-n}(y)\right],
\label{P_11-frequency}
\end{equation}
\begin{equation}
\label{P_1n-frequency}
\Pi_{1n}(\omega,y)=-\frac{e^{-y}\mathcal{E}^{2}_{n}}{2\pi l^{2}}\sum\limits_{N,N^{\prime}=0}^{\infty}
\frac{(-1)^{N+N^{\prime}}}{M_{N}M_{N^{\prime}}}\frac{N!}{(N-n)!}L^{N^{\prime}-N}_{N}
(y)L^{N-N^{\prime}}_{N^{\prime}-n}(y)\sum\limits_{\lambda,\lambda^{\prime}=\pm 1} \lambda\lambda^{\prime}\mathcal{S}(N,N',\lambda,\lambda',
\omega),
\end{equation}
where the function $S(N,N',\lambda,\lambda',\omega)$ is defined in Eq. (\ref{S-function}) and
is given in terms of digamma functions. The polarization functions (\ref{P_11-frequency}) and (\ref{P_1n-frequency}) are analytical functions of $\omega$ in the whole upper complex half plane.
The sums over the Landau levels in the two independent polarization functions are convergent
for $n \ge 3$; however, they require an ultraviolet cutoff in the case $n=1,2$ which is provided by
the band width.

The obtained expressions for polarization functions in the form of
double sums over Landau levels are most convenient to use at large magnetic fields when one can
take into account the contribution of the fewest Landau levels. To study the limit of weak magnetic
field it is better to return to Eqs. (\ref{polarization-coordinate-11}) and (\ref{polarization-coordinate-1n})
where the dependence on $N,N'$ is factorized. The weak-magnetic-field limit ($l\to\infty$) can be obtained by replacing $N\rightarrow k^2l^2/2$ with the sum turning into the integral over $k^2$
and using the asymptotic formula for the Laguerre polynomials,
\begin{equation}
L_N\left(\frac{x}{N}\right)=J_0(2\sqrt{x}),\, N\to\infty
\end{equation}
[see, Eq. (10.12.36) in Ref. \cite{Bateman}].

\section{Static and dynamical screening in clean rhombohedral graphene }
\label{screening}

\subsection{Static screening}

The static polarization functions are derived in Appendix \ref{static-lim} are given by Eqs. (\ref{static_P_11}) and (\ref{static_P_1n}) in the
general case and by Eq. (\ref{clean-static-P_11}) and (\ref{clean-static-P_1n}) in the clean case.
\begin{figure}[ht]
  \centering
  \includegraphics[scale=0.4]{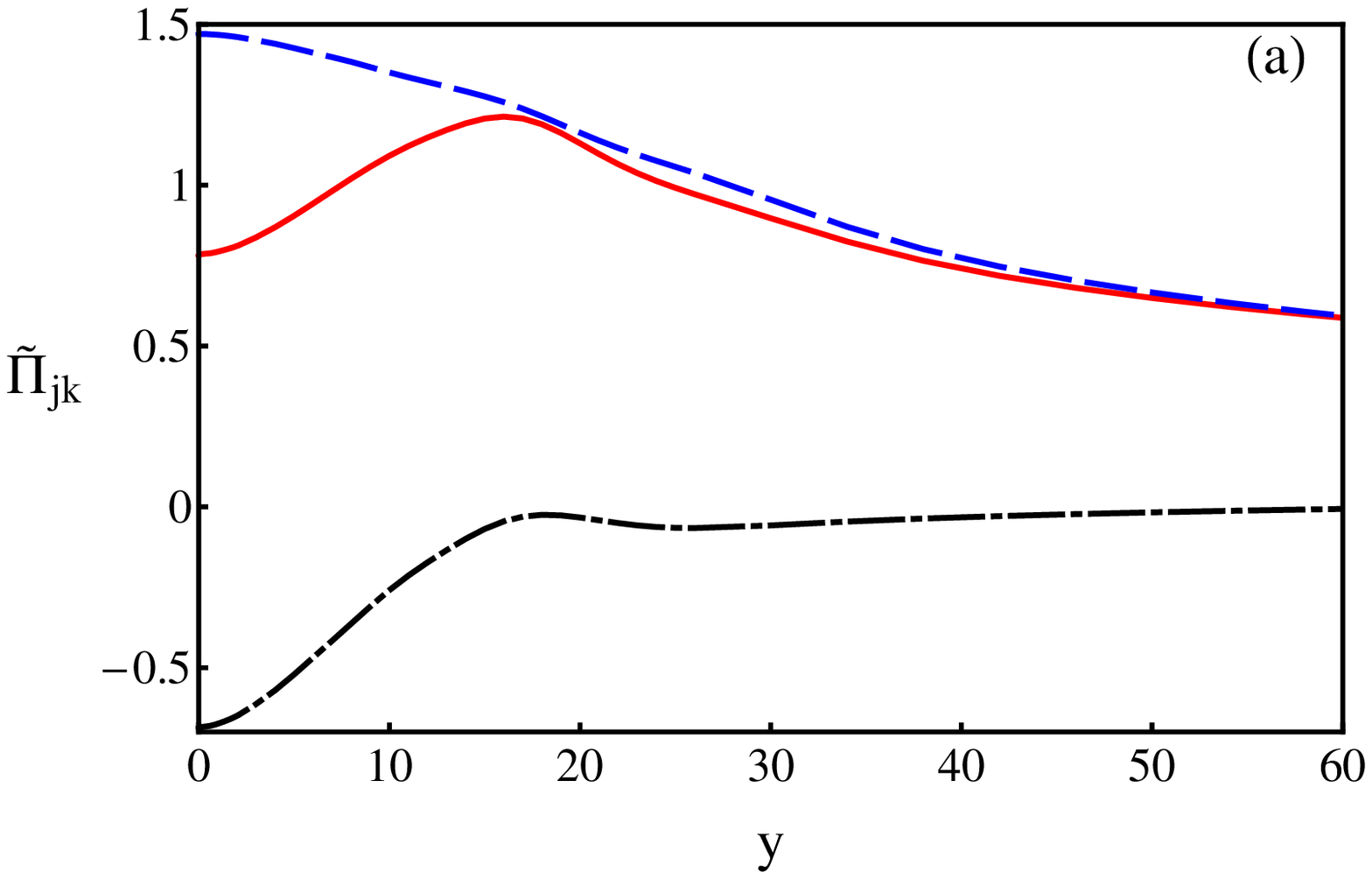}
  \includegraphics[scale=0.4]{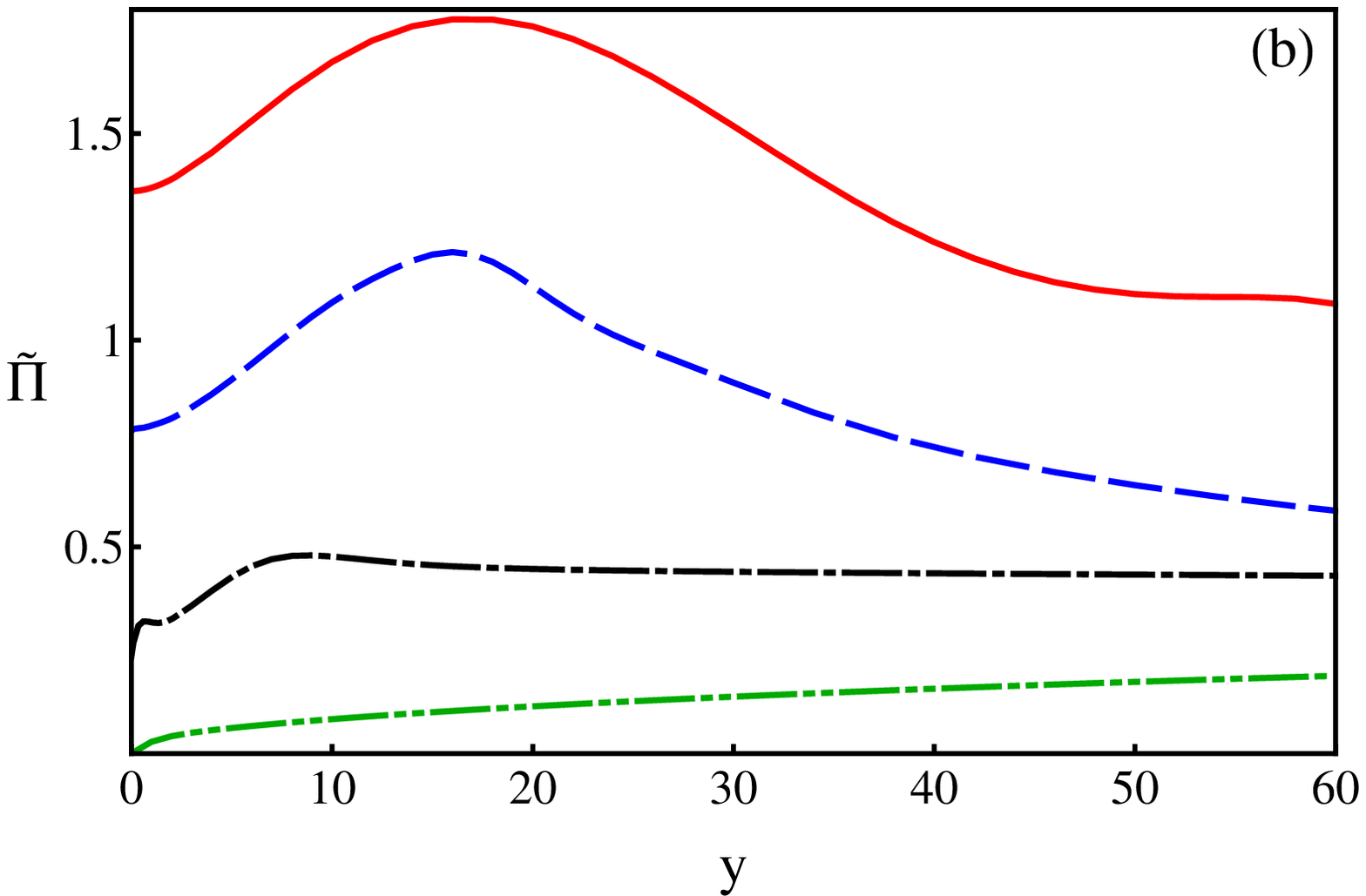}
  \caption{(Color online) The static polarization functions in gapless chiral multilayer
  graphene for $B=1$\,T, $T=2\cdot 10^{-3}\gamma_{1}$ and $\mu=0.02\gamma_{1}$. (a) Trilayer
  graphene: $\tilde{\Pi}_{11}(0,y)$ (blue dashed line), $\tilde{\Pi}_{13}(0,y)$ (black
  dash-dotted line), and $\tilde{\Pi}(0,y)$ (red solid line). (b) The function
  $\tilde{\Pi}(0,y)$ for different numbers of layers: $n=1$ (green dash-double-dotted line),
  $n=2$ (dash-dotted black line), $n=3$ (blue dashed  line), and $n=5$ (red solid  line).}
  \label{polarization-static}
\end{figure}
In order to illustrate formulas (\ref{clean-static-P_11}) and (\ref{clean-static-P_1n}) at
arbitrary momenta, we plot in Fig. \ref{polarization-static}(a) the dimensionless polarization
functions $\tilde{\Pi}_{11}(0,y)=\frac{2\hbar^2 v_{\rm F}^2}{\gamma_1}\Pi_{11}(0,y)$,
$\tilde{\Pi}_{13}(0,y)=\frac{2\hbar^2 v_{\rm F}^2}{\gamma_1}\Pi_{13}(0,y)$,
and also their sum $\tilde{\Pi}(0,y)=\tilde{\Pi}_{11}(0,y)+\tilde{\Pi}_{13}(0,y)$
in gapless  clean ABC-stacked trilayer graphene ($2\hbar^2v^2_F/\gamma_1$ is
a convenient dimensional factor which can be expressed through the Landau scale in bilayer
graphene as follows: $2\hbar^2v^2_F/\gamma_1={\cal E}_2l^2$). Clearly, at large $y$
(or $l\to\infty$), both functions tend to their zero-magnetic-field values. The function
$\tilde{\Pi}_{11}(0,y)$ monotonically decreases with $y$ while $\tilde{\Pi}_{13}(0,y)$
monotonically increases practically for all values of $y$ in the considered interval and
chosen values of temperature, chemical potential, and magnetic field.  Since $\tilde{\Pi}_{11}(0,y)$
decreases more steeply than $\tilde{\Pi}_{13}(0,y)$ increases, this leads to the appearance of
a maximum in the sum of these functions $\tilde{\Pi}(0,y)$ at some intermediate value of $y$.
In general, the behavior of these functions at intermediate values of $y$ can qualitatively
change depending on the values of parameters $lT$ and $l\mu$. Also, in panel (b) of Fig. \ref{polarization-static} we present the function $\tilde{\Pi}(0,y)$ for different numbers
of layers. As seen, the polarization functions, and consequently the screening effects,
increase with the number of layers $n$.

\begin{figure}[ht]
  \centering
  \includegraphics[scale=0.4]{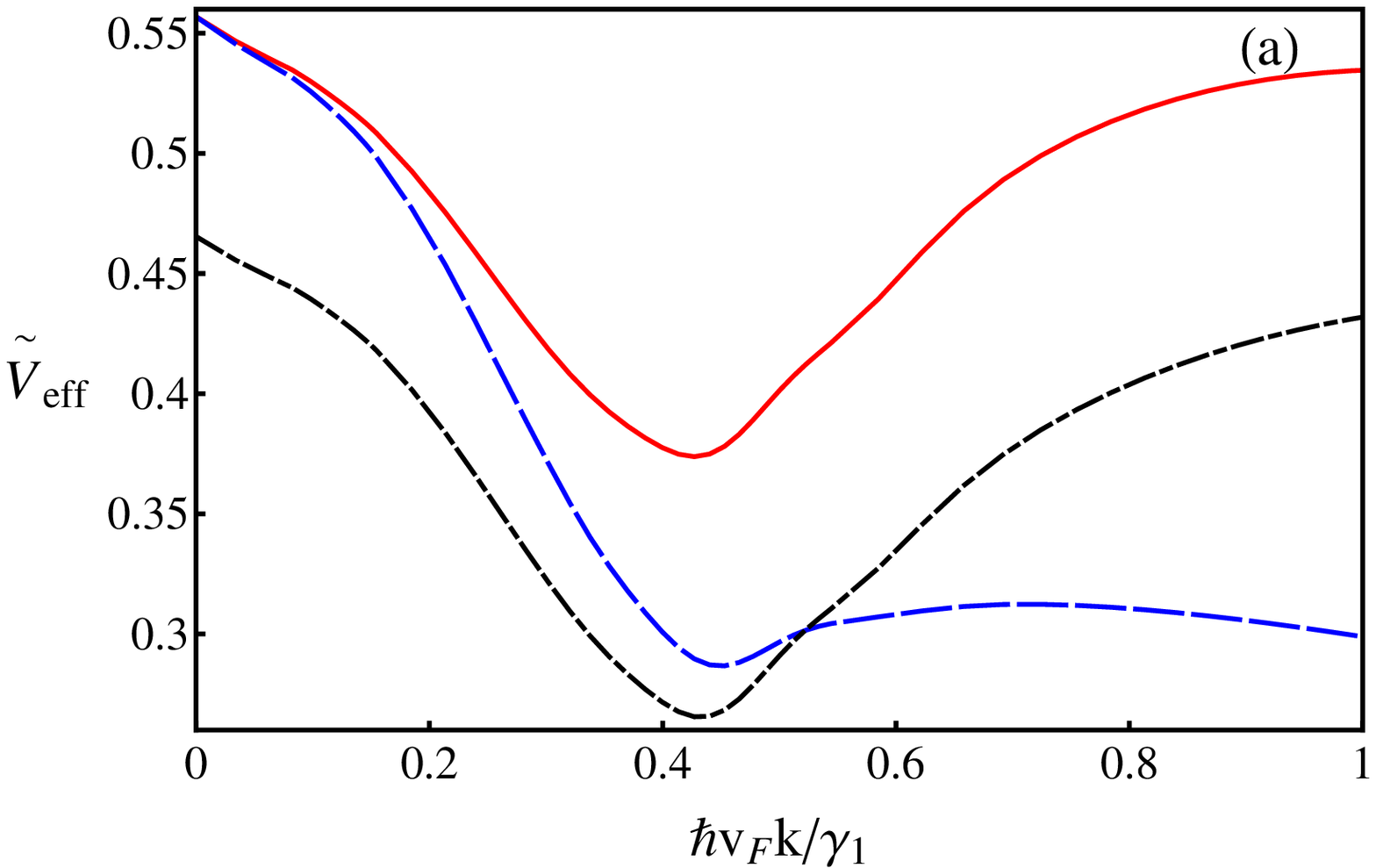}
  \includegraphics[scale=0.4]{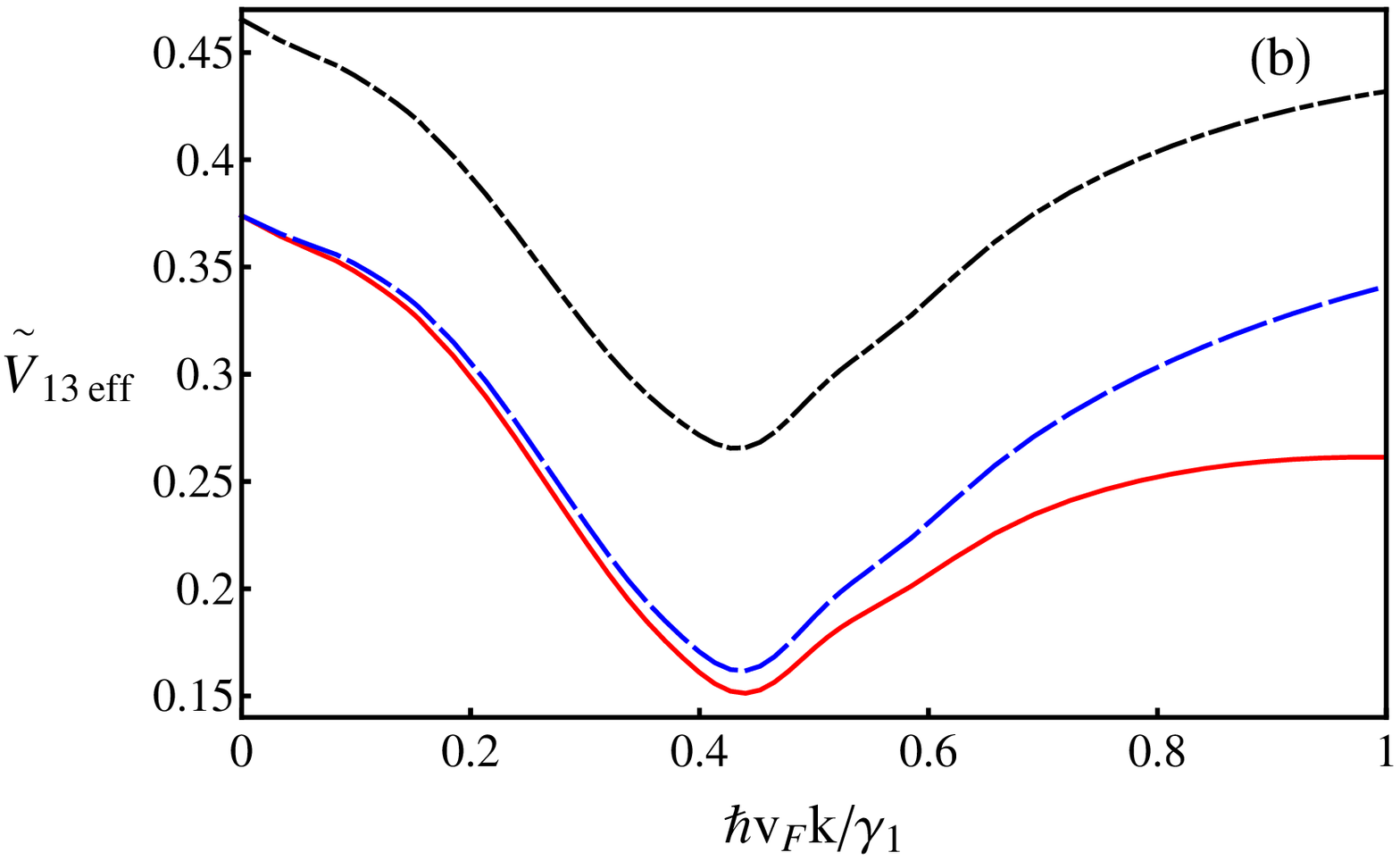}
  \caption{(Color online) The dimensionless effective static intralayer, $\tilde{V}_{eff}$,
  and interlayer, $\tilde{V}_{13\,eff}$, interactions as functions of $\hbar v_F k/\gamma_1$
  in gapless trilayer graphene for $B=1$T, $T=0.002\gamma_{1}$ and $\mu=0.02\gamma_{1}$ (the corresponding filling factor $\nu=2.7$).
  (a) Red solid line shows Eq. (\ref{interaction-effective}), blue dashed
  line shows Eq. (\ref{interaction-effective_simp2}), and black dash-dotted line shows
  Eq. (\ref{interaction-effective_simp3}). (b) Red solid line shows Eq. (\ref{interaction-ND}),
  blue dashed line shows Eq. (\ref{interaction-ND_simp2}), and black dash-dotted line shows
  Eq. (\ref{interaction-effective_simp3}).}
  \label{V}
\end{figure}

We plot in Fig. \ref{V} the dimensionless [multiplied by the factor $\kappa \gamma_{1}/(2\pi e^2
\hbar v_{\rm F})$] effective interactions given
by Eqs. (\ref{interaction-effective}) and (\ref{interaction-ND}) and approximate expressions (\ref{interaction-effective_simp2}), (\ref{interaction-ND_simp2}), and
(\ref{interaction-effective_simp3}) in the ABC-stacked trilayer graphene for $\omega=0$ and
$\kappa=5$: $\tilde{V}_{eff}(0,k)=V_{eff}(0,k)\kappa \gamma_{1}/(2\pi e^2 \hbar v_{\rm F})$,
$\tilde{V}_{13\,eff}(0,k)=V_{13\,eff}(0,k)\kappa \gamma_{1}/(2\pi e^2 \hbar v_{\rm F})$.
As expected, ${V}^{IR}_{\mbox{\scriptsize eff}}$ and ${V}^{IR}_{1n\,
\mbox{\scriptsize eff}}$ given by Eqs. (\ref{interaction-effective_simp2}) and
(\ref{interaction-ND_simp2}) and plotted as blue dashed lines in Fig. \ref{V},
excellently match for small momenta the effective interactions (\ref{interaction-effective})
and (\ref{interaction-ND}) plotted as red solid lines. [Note that the appearance of a minimum in
the screened Coulomb interactions is connected with the presence of a maximum in the
polarization function $\tilde{\Pi}(0,y)$.] However, the corresponding curves deviate
significantly for large momenta (especially so for ${V}^{IR}_{\mbox{\scriptsize eff}}$
and ${V}_{\mbox{\scriptsize eff}}$). The approximate effective interactions, given by Eq.  (\ref{interaction-effective_simp3}) and plotted as black dash-dotted lines in Fig. \ref{V},
reproduce well the behavior of the effective interactions (\ref{interaction-effective})
and (\ref{interaction-ND}) for all $k$. The two curves are only shifted with respect to one another.
 We checked that this is true for $n\le8$; still
the agreement worsens (especially so at large $k$) as $n$ grows.  Moreover, the discrepancies
between the corresponding dependencies can be practically completely eliminated by
adjusting the dielectric constants in ${V}^{\rm appr}_{\mbox{\scriptsize eff}}$ and ${V}^{\rm appr}_{1n\,\mbox{\scriptsize eff}}$. Since the approximate effective interactions (\ref{interaction-effective_simp3}) depend only on $\Pi=\Pi_{11}+\Pi_{1n}$,
further in this section we will analyze the polarization function $\Pi$.
\begin{figure}[ht]
  \centering
  \includegraphics[scale=0.4]{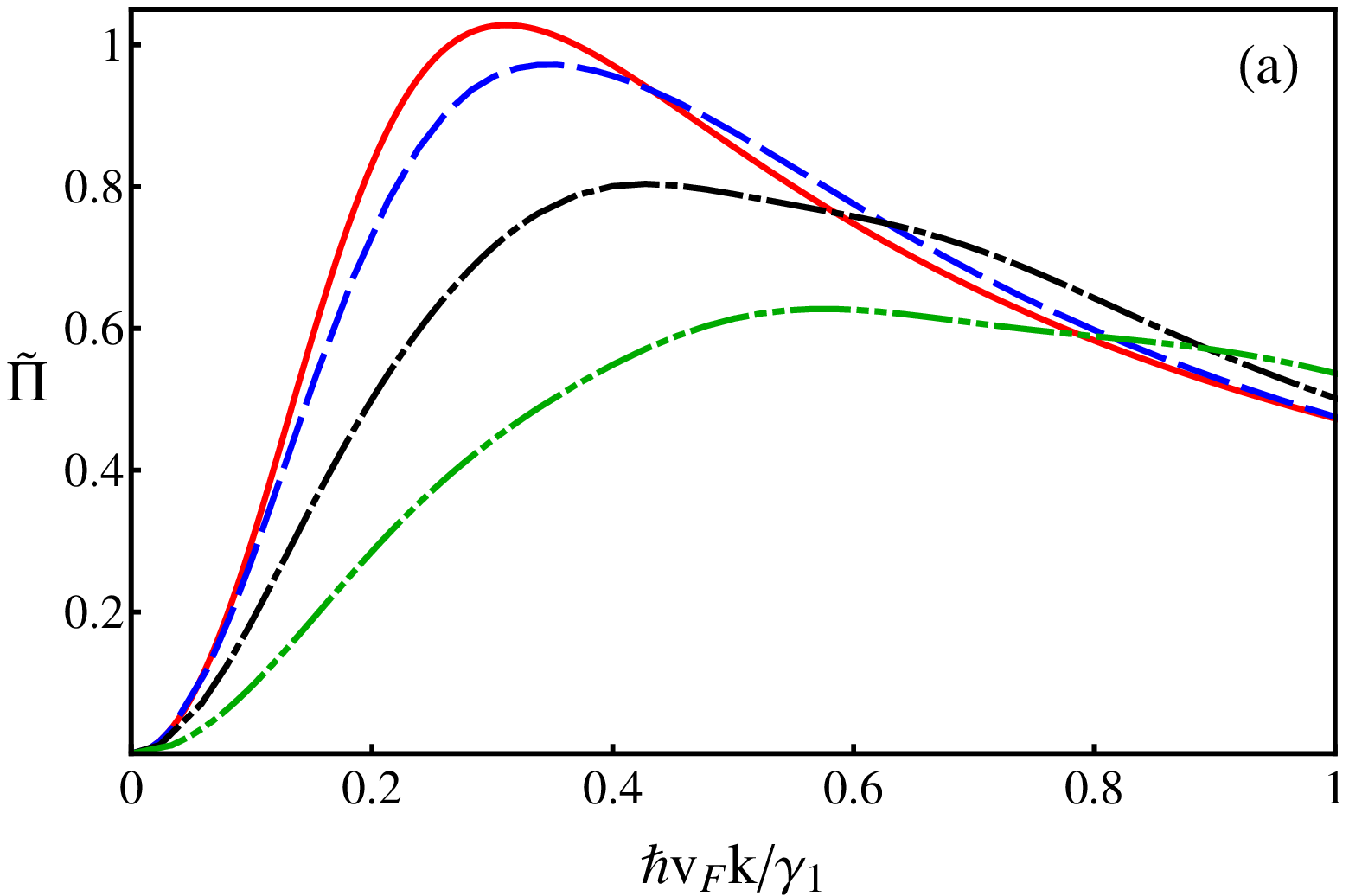}
  \includegraphics[scale=0.41]{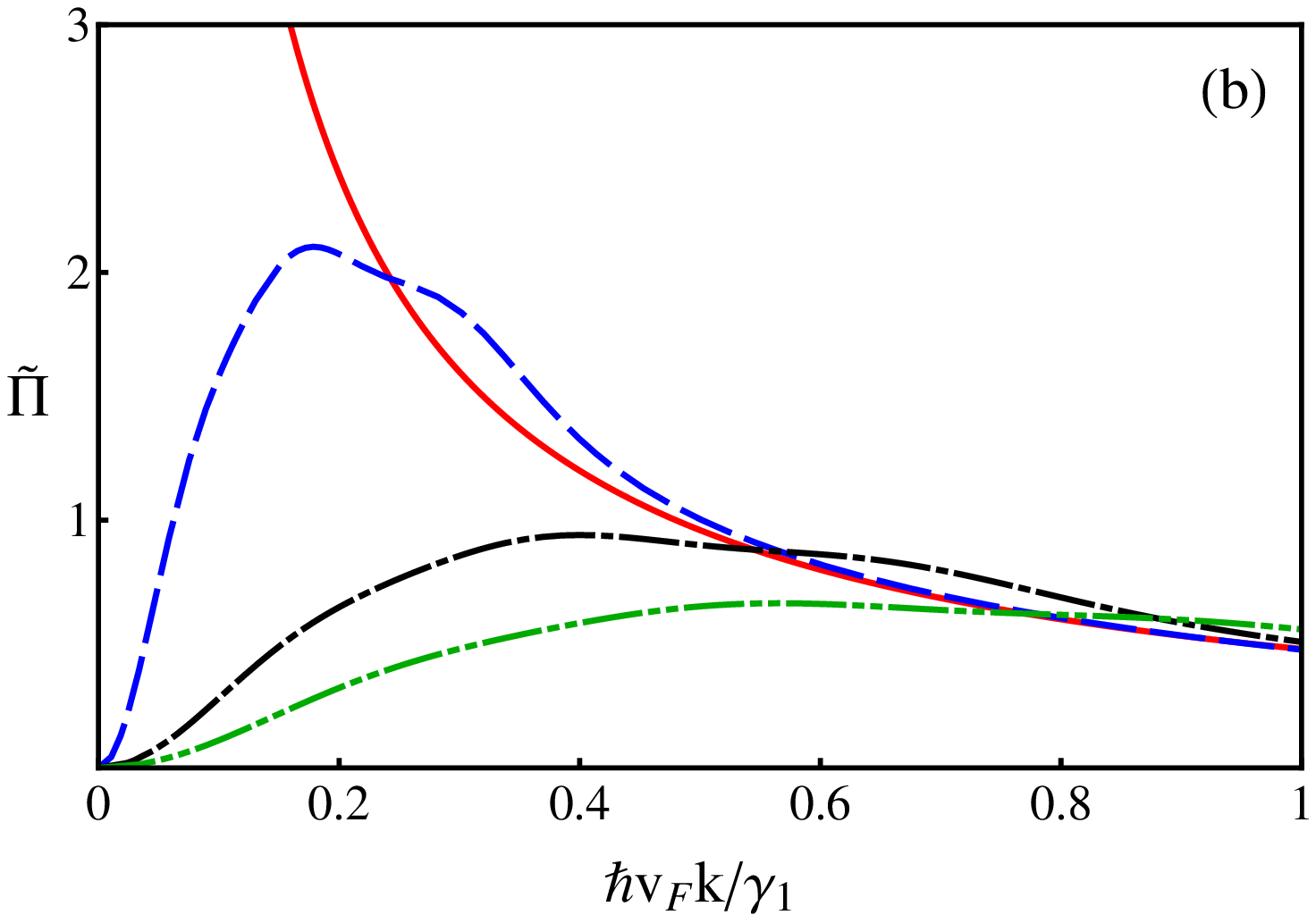}
  \caption{(Color online) The dimensionless static polarization function $\tilde{\Pi}(0,k)$ in
  gapped and gapless trilayer graphene, for $T=\mu=0$, and different values of the
  magnetic field: red solid lines are for $B=0$, blue dashed lines are for $B=1$\,T, black dash-dotted lines
  are for $B=5$\,T, green dash-double-dotted lines are for $B=10$\,T. (a) $\Delta=0.01\gamma_1$. (b)
  $\Delta=0$.}
  \label{pol-B0andnonzeroB}
\end{figure}

In order to see how the magnetic field affects the screening, we plot in Fig. \ref{pol-B0andnonzeroB} the static
dimensionless polarization function $\tilde{\Pi}(0,k)$ as a function of the wave vector
$k=|\mathbf{k}|$ for different values of magnetic field  in gapped
and gapless  trilayer graphene. Clearly, the magnetic field suppresses the screening.
According to Fig. \ref{pol-B0andnonzeroB}(a), the larger is the magnetic field,
the stronger is the suppression (compare
blue, green, and black lines). The role of magnetic field is especially dramatic in gapless graphene.
The magnetic field qualitatively changes the behavior of the polarization functions at small
wavevectors $|\mathbf{k}| \le l^{-1}$. Indeed, while the static polarization function $\Pi(0,\mathbf{k})$ diverges as $\mathbf{k} \to 0$ for $n \ge 3$ in the gapless case without
magnetic field [see, red line in the panel Fig. \ref{pol-B0andnonzeroB}(b)], it is no longer divergent in the presence of a magnetic field  (see, blue, green, and black lines).

The asymptotical behavior of the screened Coulomb potential at small and large distances is
determined by the asymptotics of the polarization function $\Pi(0,\mathbf{k})=\Pi_{11}(0,\mathbf{k})+\Pi_{1n}(0,\mathbf{k})$
at large and small wave vectors, respectively. At small wave vector values ($\mathbf{k}\to 0$),
the behavior of $\Pi(0,\mathbf{k})$ can be found from Eqs. (\ref{static_P_11}) and
(\ref{static_P_1n}) by taking into account the values of Laguerre polynomials at zero:
$L^m_n(0)=(m+n)!/(m!n!)\,$, $(L^m_n)'(0)=-(m+n)!/[(m+1)!(n-1)!]$. By using
${\cal E}^2_nN!/(N-n)!=M_N^2-\Delta^2$, we get
\begin{equation}
\Pi(0,\mathbf{k})\simeq a+b\mathbf{k}^2,
\label{staticPi-k-to0}
\end{equation}
where the coefficients $a$ and $b$ are expressed in terms of the digamma function and its derivative
\begin{eqnarray}
a=\Pi(0,0)=\frac{1}{2\pi^3 l^{2} T}\sum\limits_{N=0}^{\infty}{\sum\limits_{\lambda=\pm 1}}^{\prime}
\Re e\psi^{\prime}\left(\frac{1}{2}+\frac{\Gamma_{N}}{2\pi T}+\frac{\mu-\lambda M_{N}}
{2\pi i T}\right),
\label{Pi(0,0)}
\end{eqnarray}
\begin{eqnarray}
b&=&-\frac{1}{4\pi^3 T}\sum\limits_{N=0}^{\infty}{\sum\limits_{\lambda=\pm 1}}^{\prime}\left[2N+1-n\left(1+\lambda\frac{\Delta}
{M_{N}}\right)\right]\Re e\psi^{\prime}\left(\frac{1}{2}+\frac{\Gamma_{N}}{2\pi T}+
\frac{\mu-\lambda M_{N}}{2\pi i T}\right)\nonumber\\
&+&\frac{1}{4\pi^2}\sum\limits_{N=0}^{\infty}{\sum\limits_{\lambda,\lambda^{\prime}=
\pm 1}}^{\prime \prime}\Im m\left\{\frac{\psi\left(\frac{1}{2}+\frac{\Gamma_{N}}{2\pi T}+\frac{\mu-\lambda M_{N}}{2\pi i
T}\right)-\psi\left(\frac{1}{2}+\frac{\Gamma_{N+1}}
{2\pi T}+\frac{\mu-\lambda^{\prime} M_{N+1}}{2\pi i T}\right)}{\lambda M_{N}-\lambda^{\prime}
M_{N+1} -i\left(\Gamma_{N}-\Gamma_{N+1}\right)}\right\}\nonumber\\
&\times&\left[2(N+1)\left(1+\lambda\lambda^{\prime}\frac{M_{N}}{M_{N+1}}\right)
-n\left(1+\lambda\frac{\Delta}{M_{N}}\right)\left(1+\lambda^{\prime}\frac{\Delta}{M_{N+1}}
\right)\right].
\label{coeff-b}
\end{eqnarray}
Here the notation $\sum'$ means that only the term with  $\lambda=-1$ is retained in the sum
if $N<n$, and $\sum''$ means that only the terms with $\lambda=-1$ (if $N<n$) and $\lambda'=-1$
(if $N<n-1$) are retained. One can see that the parameter $a$ receives the contribution only from intra-Landau-level ($N\leftrightarrow N$) transitions while the parameter $b$ contains also the contributions from the transitions between adjacent Landau levels $N\leftrightarrow  N\pm1$.

The polarization function at zero frequency and momentum (\ref{Pi(0,0)}) obeys
the following relation \cite{Davies}:
\begin{equation}
\Pi(0,0)=\int\frac{d\epsilon D(\epsilon)}{4T\cosh^2\left(\frac{\epsilon-\mu}{2T}\right)},
\end{equation}
where $D(\epsilon)$ is the density of states (DOS) in multilayer graphene with impurities in
a magnetic field,
\begin{equation}
D(\epsilon)=\frac{1}{\pi^2l^2}\sum\limits_{N=0}^{\infty}{\sum\limits_{\lambda=\pm 1}}^{\prime}
\frac{\Gamma_N}{(\epsilon-\lambda M_N)^2+\Gamma_N^2}.
\end{equation}
Obviously, $\Pi(0,0)$ is an oscillating function of chemical potential and magnetic field.
At zero temperature and finite scattering rate it is simply equal to the DOS at the Fermi surface
$\Pi(0,0)=D(\mu)$. We would like to remind the reader that the screening of the Coulomb potential at large
distances is determined by the magnitude of the Thomas-Fermi wave vector $k_F=(4\pi e^2/\kappa)\Pi(0,0)$. When the Fermi level lies between Landau levels (that corresponds to an integer filling)
$\Pi(0,0)$ reaches a minimum (zero for $T=0, \Gamma_N=0$) and the screening is minimal. On the
other hand, if the Fermi level is inside a Landau level, then the screening is maximal. The
corresponding screening properties of electronic environment on an isolated charged impurity
in a magnetic field were recently observed experimentally in monolayer graphene in
Ref. \cite{Andrei}. It was demonstrated that, in the presence of a magnetic field, the strength
of the impurity can be tuned by controlling the occupation of Landau levels with a gate voltage.
It would be interesting to repeat such experiments in multilayer graphene.
\begin{figure}[ht]
  \centering
  \includegraphics[scale=0.425]{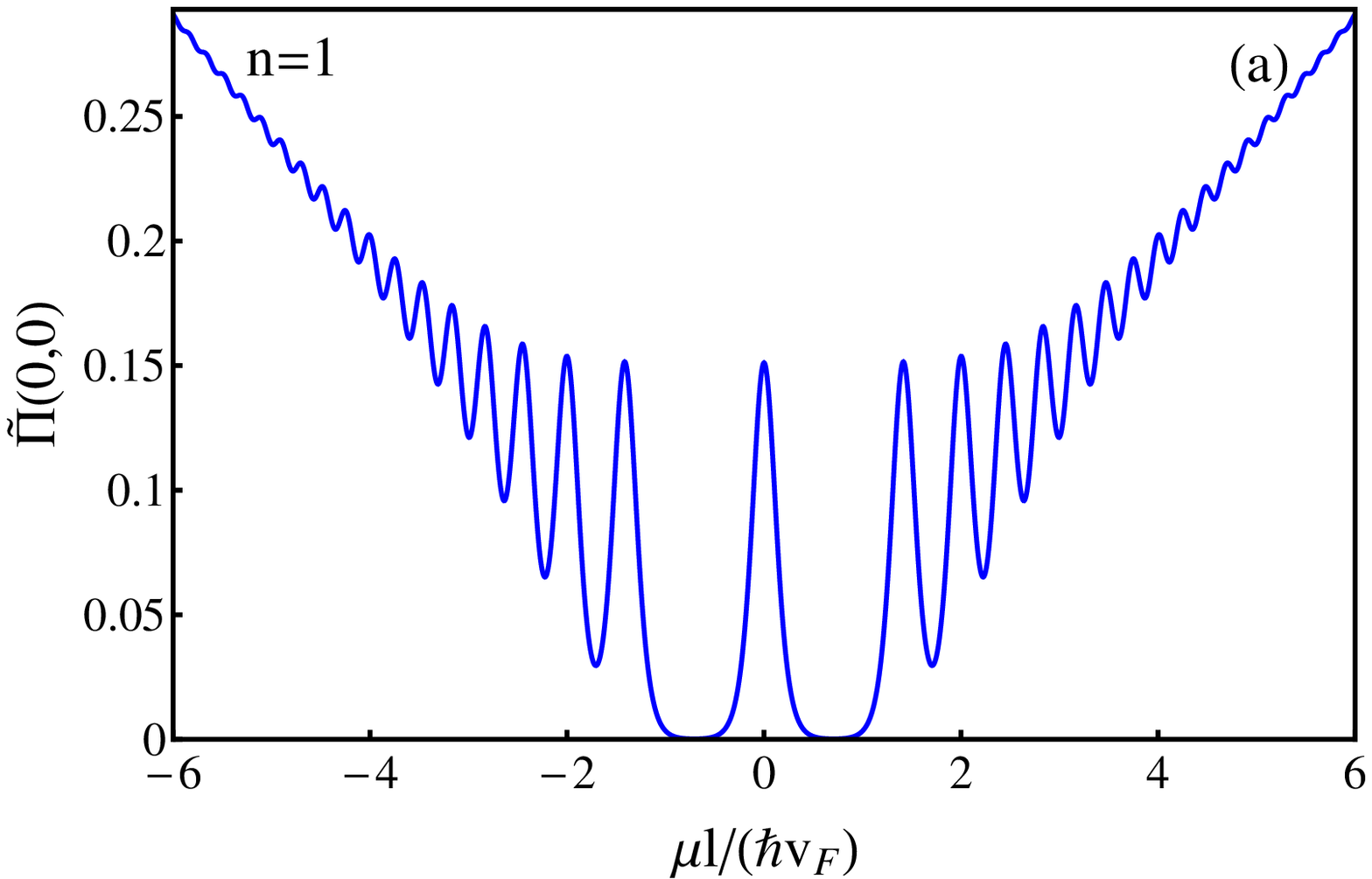}
  \includegraphics[scale=0.425]{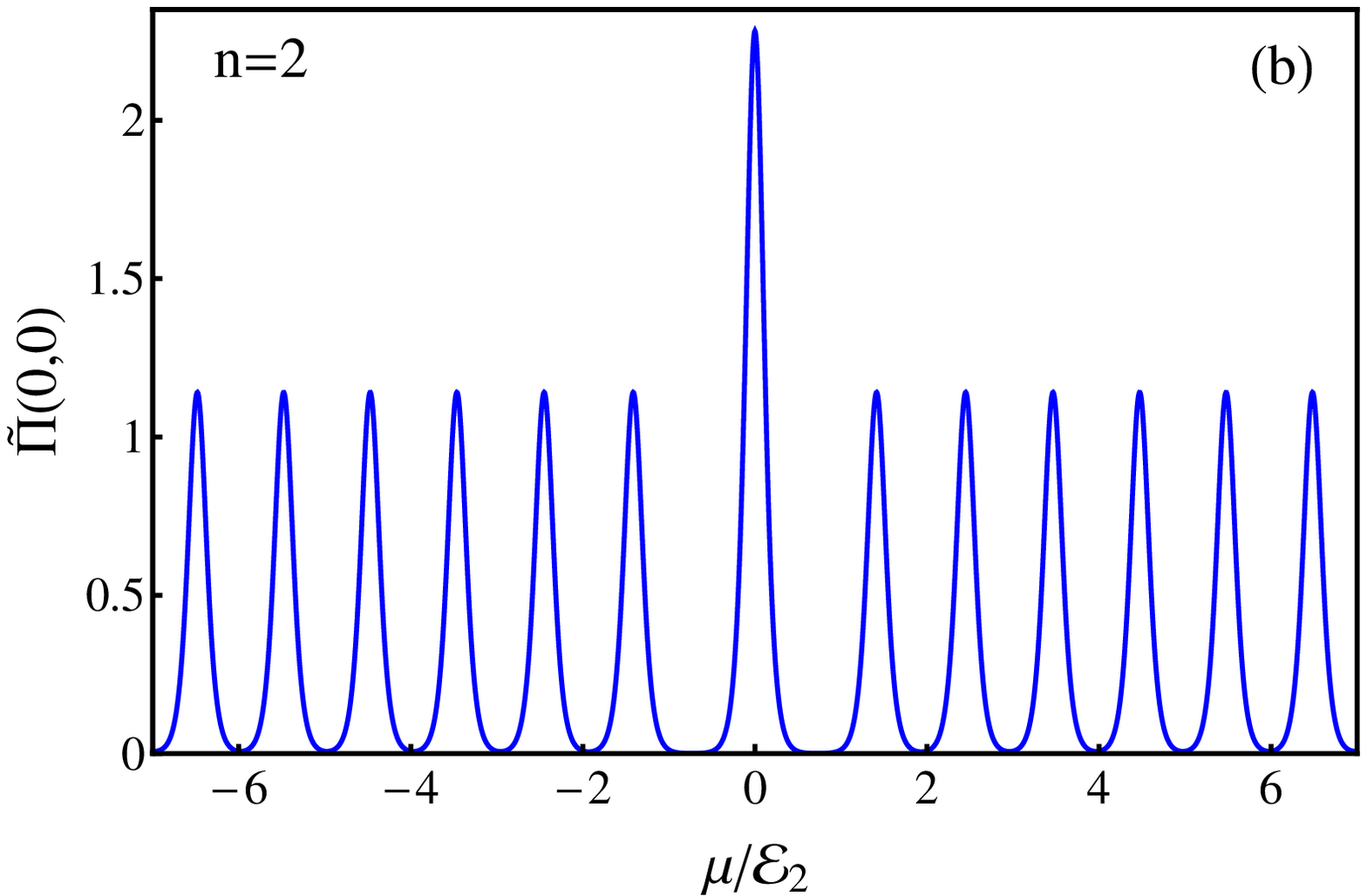}
  \includegraphics[scale=0.425]{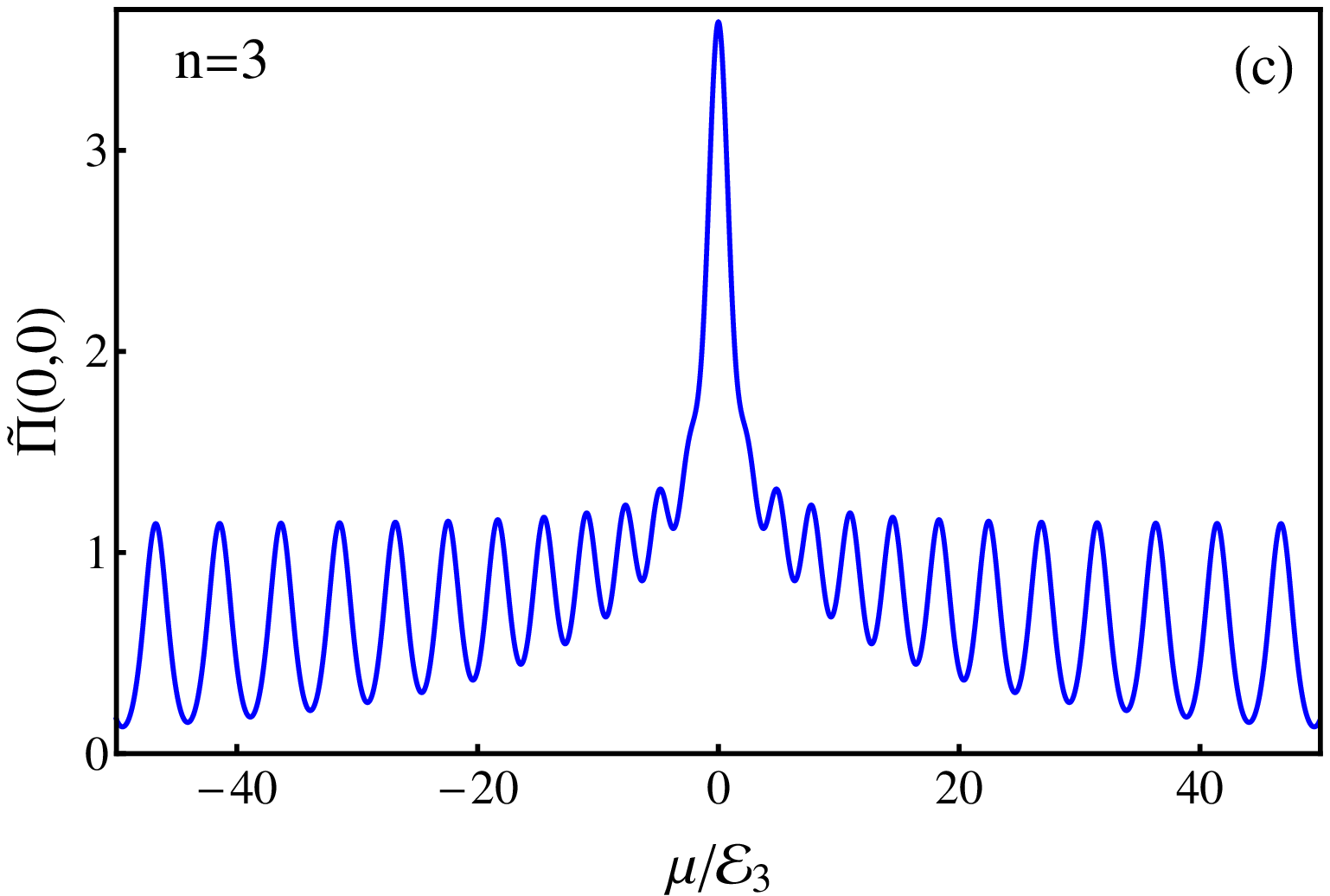}
  \includegraphics[scale=0.425]{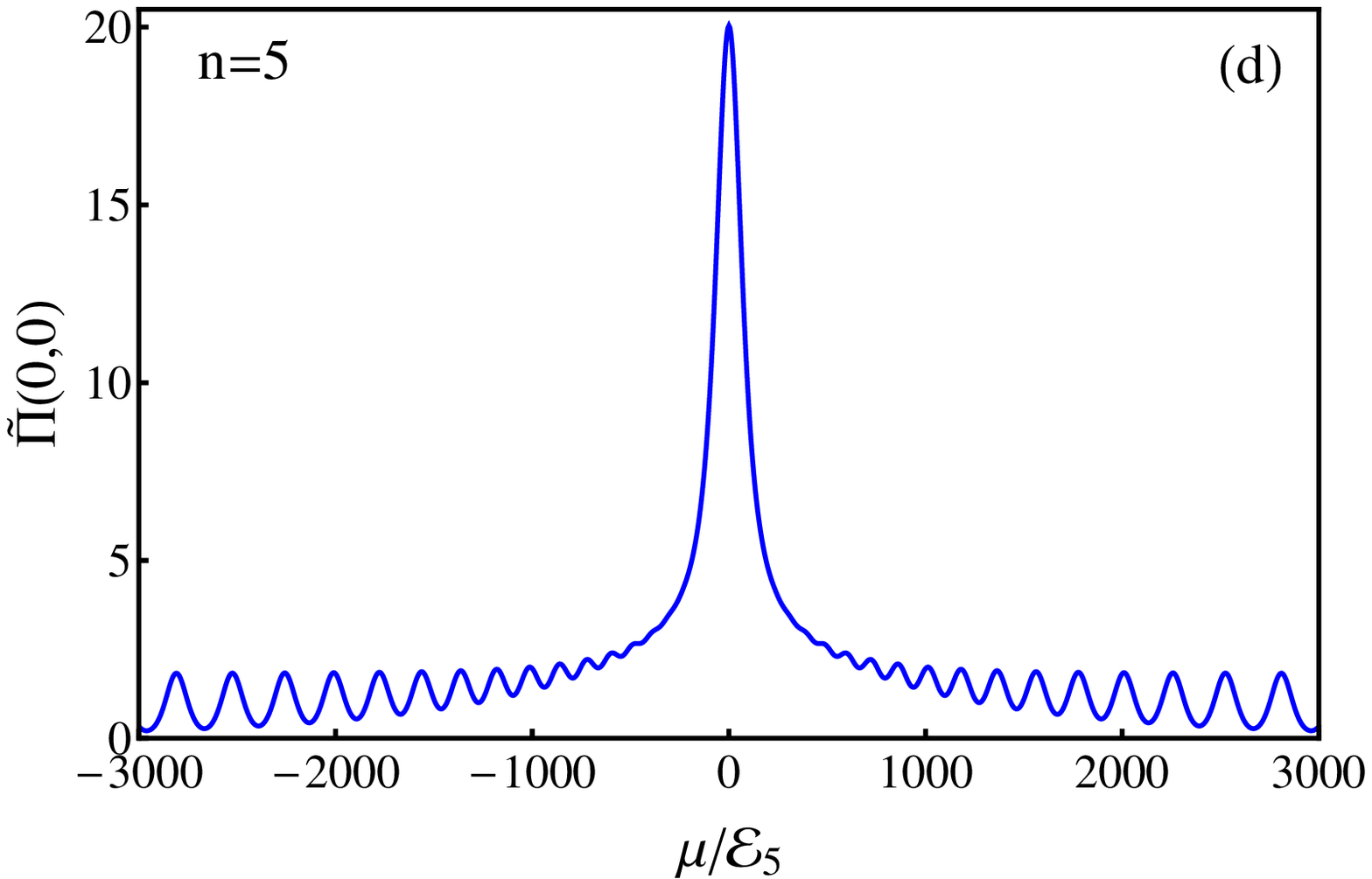}
  \caption{(Color online) $\tilde{\Pi}(0,0)$ as a function of $\mu$ in the gapless case $\Delta=0$,
  and magnetic field $B=1$\,T.
   (a) $n=1$ ($T=5\cdot 10^{-3}\gamma_{1}$), (b) $n=2$ ($T=8\cdot 10^{-4}\gamma_{1}$),
   (c) $n=3$ ($T=8\cdot 10^{-4}\gamma_{1}$), and (d) $n=5$ ($T=5\cdot 10^{-4}\gamma_{1}$).}
 \label{figsPi00Delta0}
\end{figure}

\begin{figure}[ht]
  \centering
  \includegraphics[scale=0.425]{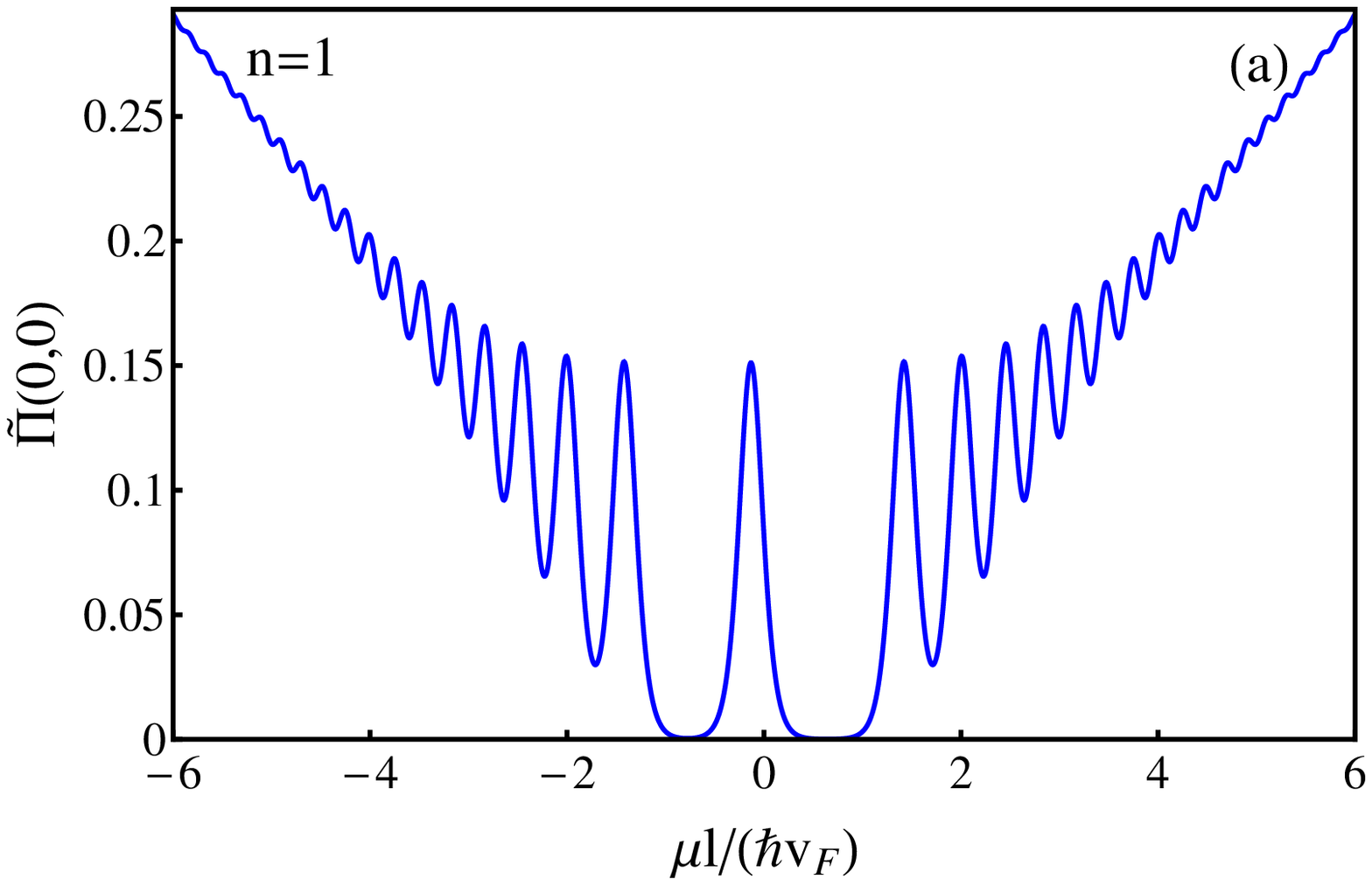}
  \includegraphics[scale=0.425]{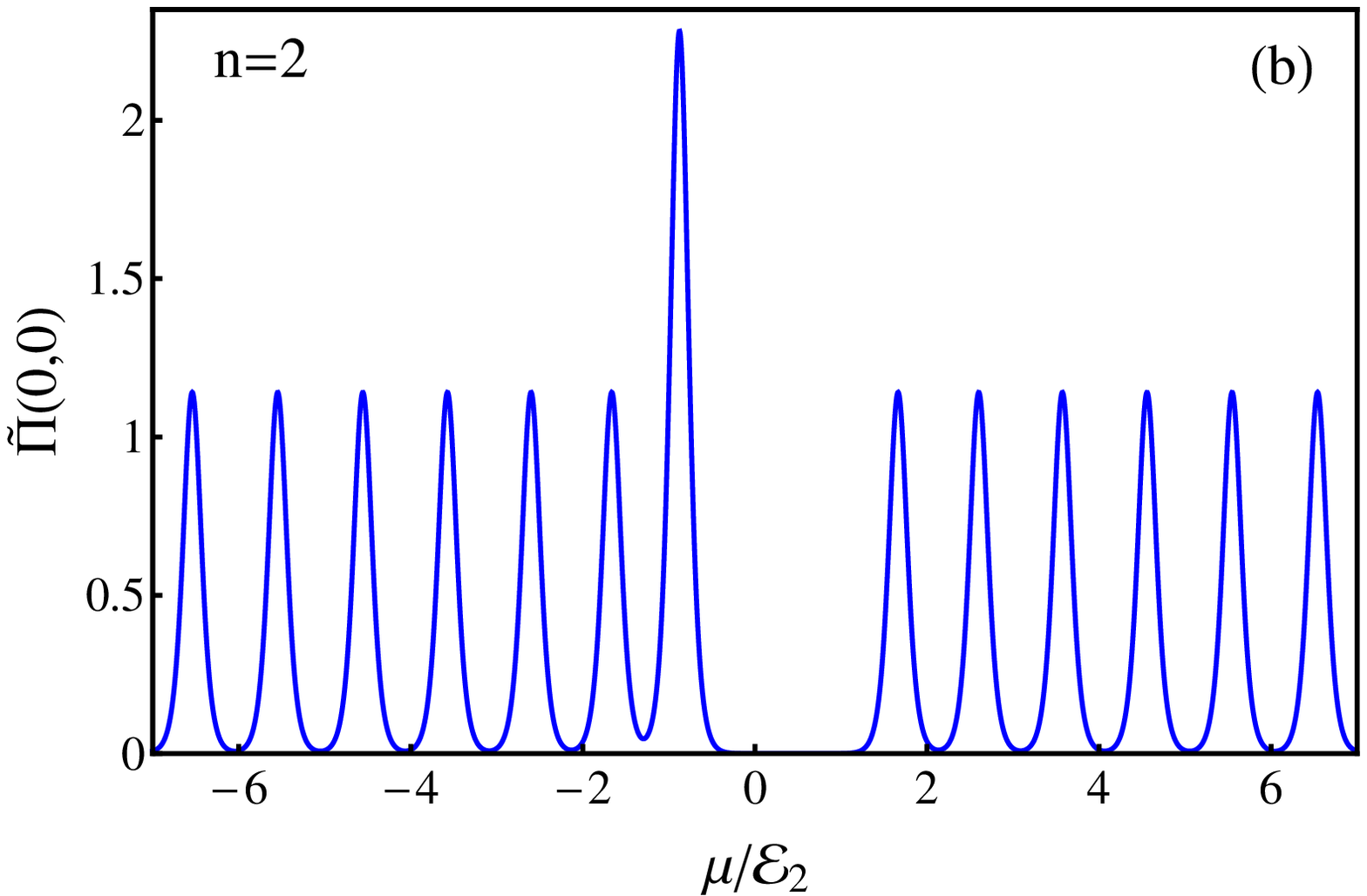}
  \includegraphics[scale=0.425]{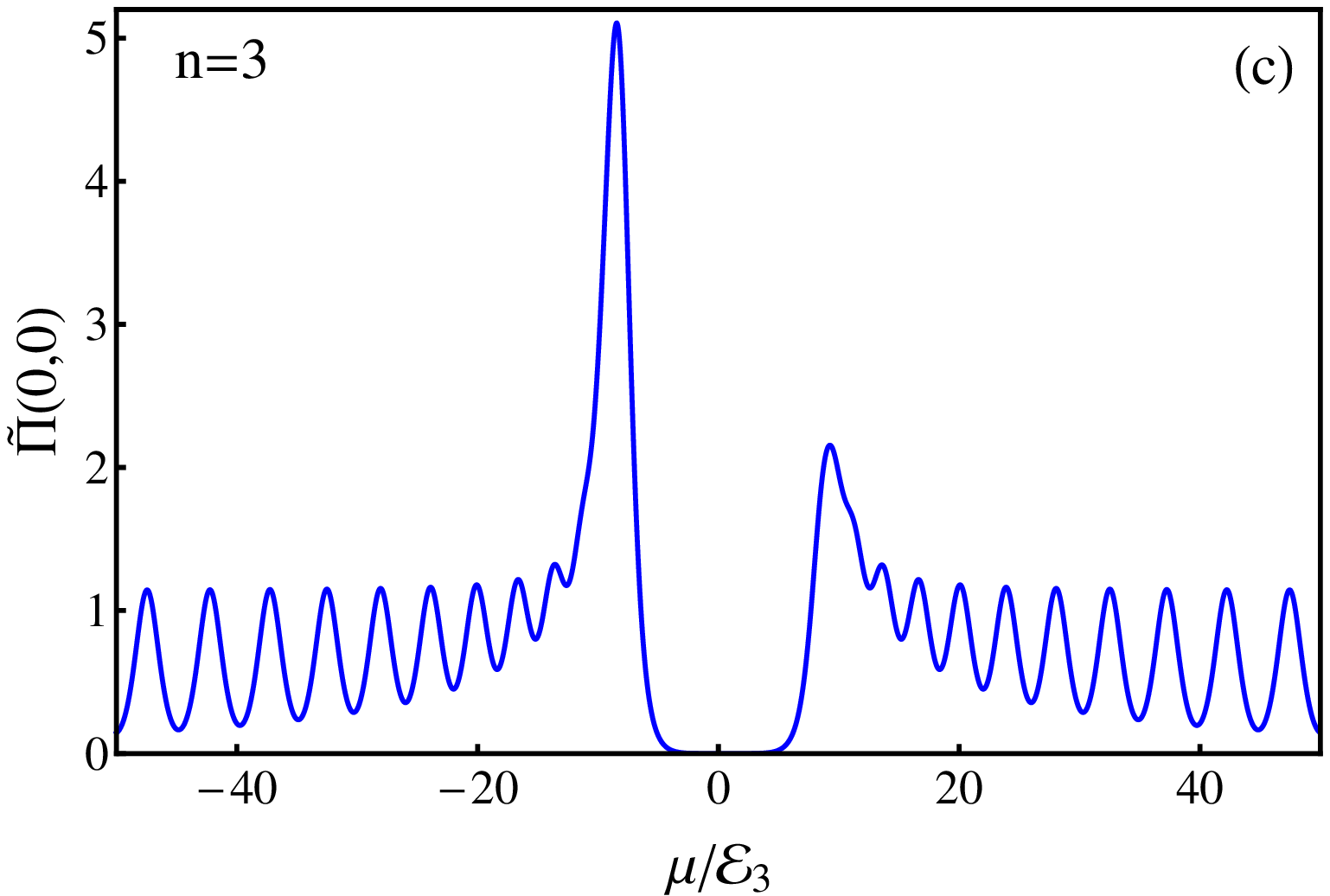}
  \includegraphics[scale=0.425]{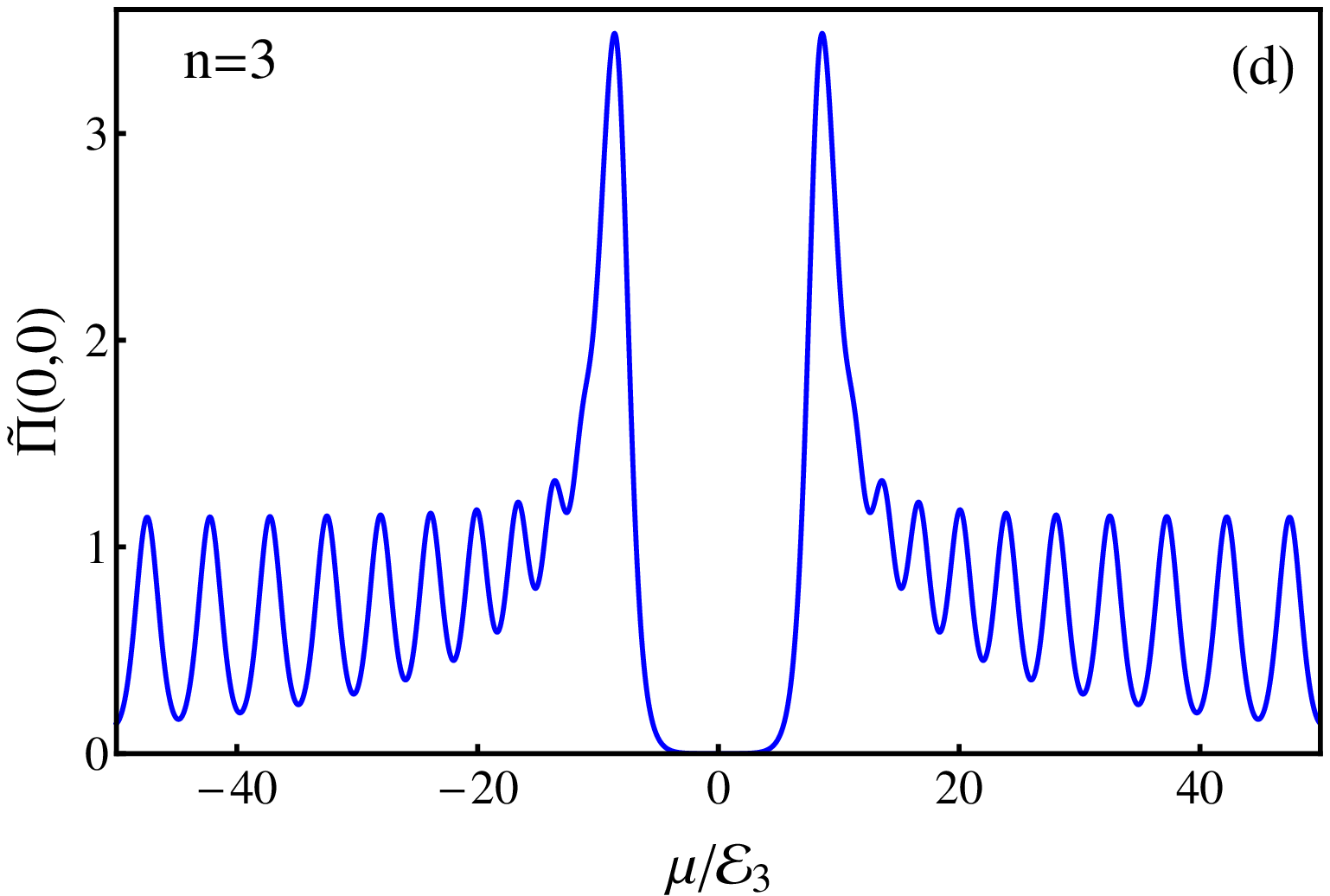}
  \caption{(Color online) $\tilde{\Pi}(0,0)$ as a function of $\mu$ for the QAH state gap
  $\Delta_{\xi s}=\Delta =0.01\gamma_{1}$ and the magnetic field $B=1$T: (a)
  $n=1$ ($T=5\cdot 10^{-3}\gamma_{1}$), (b) $n=2$ ($T=8\cdot 10^{-4}\gamma_{1}$),
   (c) $n=3$ ($T=8\cdot 10^{-4}\gamma_{1}$), and (d) for the LAF state gap
   $\Delta_{\xi s}=\xi s\Delta$, $\Delta=0.01\gamma_{1}$, $n=3$ ($T=8\cdot 10^{-4}\gamma_{1}$).}
   \label{figsPi00Delta_finite}
\end{figure}
The static polarization functions  Eqs. (\ref{static_P_11}) and (\ref{static_P_1n}) simplify in
the case of zero width Landau levels $\Gamma_N=0$ when they are described by
Eqs. (\ref{clean-static-P_11}) and (\ref{clean-static-P_1n}). The quantity $\Pi(0,0)$, which
is related to the Thomas-Fermi screening vector, is given in the clean ABC-stacked multilayer
graphene by the expression
\begin{equation}
\Pi(0,0)=\frac{1}{4\pi l^2 T}\sum\limits_{N=n}^{\infty}\sum\limits_{\lambda=\pm 1}
\frac{1}{{\rm ch}^2\left(\frac{\mu-\lambda M_{N}}{2T}\right)}+\frac{1}{4\pi l^2 T}
\frac{n}{{\rm ch}^2\left(\frac{\mu+\Delta}{2T}\right)}.
\end{equation}
The weak-magnetic-field limit ($l\to\infty$) of the above expression can be obtained by
replacing $N\rightarrow k^2l^2/2$, with the sum turning into the integral
\begin{equation}
\Pi(0,0)=\frac{1}{4\pi Tn(a_n)^{2/n}}\int_{-\infty}^\infty\frac{d\epsilon|\epsilon|(\epsilon^2
-\Delta^2)^{1/n -1}\theta(\epsilon^2-\Delta^2)}{\cosh^2\left(\frac{\epsilon-\mu}{2 T}\right)},
\end{equation}
which at $T=0$ gives the density of states of chiral multilayer graphene at the Fermi surface
in the absence of the magnetic field
\begin{equation}
\Pi_{T=0}(0,0)=D(\mu)=\frac{|\mu|}{n \pi\hbar^2 v_{F}^{2}}\left(\frac{\gamma_{1}^{2}}{\mu^2-\Delta^2}
\right)^{1-\frac{1}{n}}\Theta(\mu^2-\Delta^2).
\label{free-DOS}
\end{equation}
In Fig. \ref{figsPi00Delta0} we show the dependence of the dimensionless quantity $\tilde{\Pi}(0,0)=(2\hbar^2v_F^2/\gamma_1)\Pi(0,0)$ for
$n = 1, 2, 3,5$ in gapless rhombohedral graphene. Comparing the behavior of the DOS as
the number of layers $n$ increases one can see that the LLL contribution is enhanced because
Landau levels are getting denser near zero energy for $n>2$ (notice the change of scale of
the vertical axis). Furthermore, the DOS envelope
function itself reflects the behavior of the zero-field DOS given by Eq. (\ref{free-DOS}):
for large $\mu$ it grows for $n=1$, becomes constant for $n=2$, and decreases for $n\ge3$.
The strength of the peak corresponding to the zero energy LL also increases as the number
of layers grows that reflects the degeneracy of this level. These results for the DOS agree
with those discussed in Refs. \cite{Gusynin-Pyatkovskiy} ($n=1$) and \cite{FNT2014} ($n=2,3$).

In Fig. \ref{figsPi00Delta_finite} we plot also the quantity $\tilde{\Pi}(0,0)$ as a function of
$\mu$ in the gapped case $\Delta_{\xi s} =\Delta$ which corresponds to the quantum anomalous Hall
(QAH) state \cite{QAH,LAF} with broken time-reversal symmetry. One can observe an asymmetry of
this function with respect to the change $\mu\to-\mu$ which is a consequence of the asymmetry
of the lowest Landau level for the QAH state: there is a state with the energy $E=-\Delta$ and
no a state with $E=\Delta$. For the layer antiferromagnetic  (LAF) state \cite{LAF,Kharitonov,Roy}
with the gap $\Delta_{\xi s}=\xi s\Delta$ which also breaks time-reversal symmetry,
the corresponding symmetry of $\tilde{\Pi}(0,0)$ is restored [see, Fig. \ref{figsPi00Delta_finite}(d) for $n=3$]. We note that the recent experimental data
in Ref. \cite{Gillgren} suggest that the LAF state is the ground state of rhombohedral trilayer
graphene in the absence of external fields.

\subsection{Dynamical screening}

\begin{figure}[ht]
  \centering
  \includegraphics[scale=0.4]{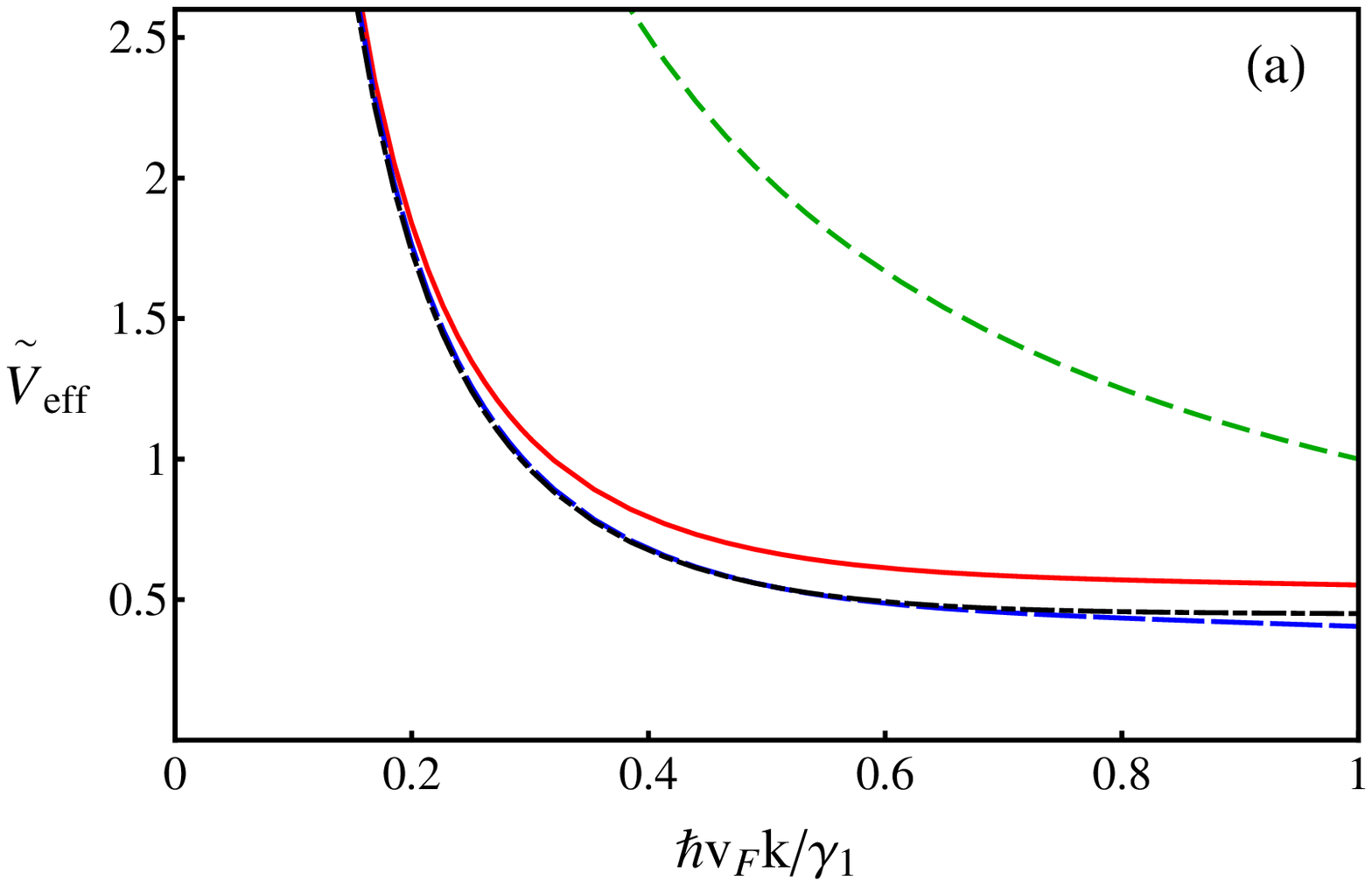}
  \includegraphics[scale=0.4]{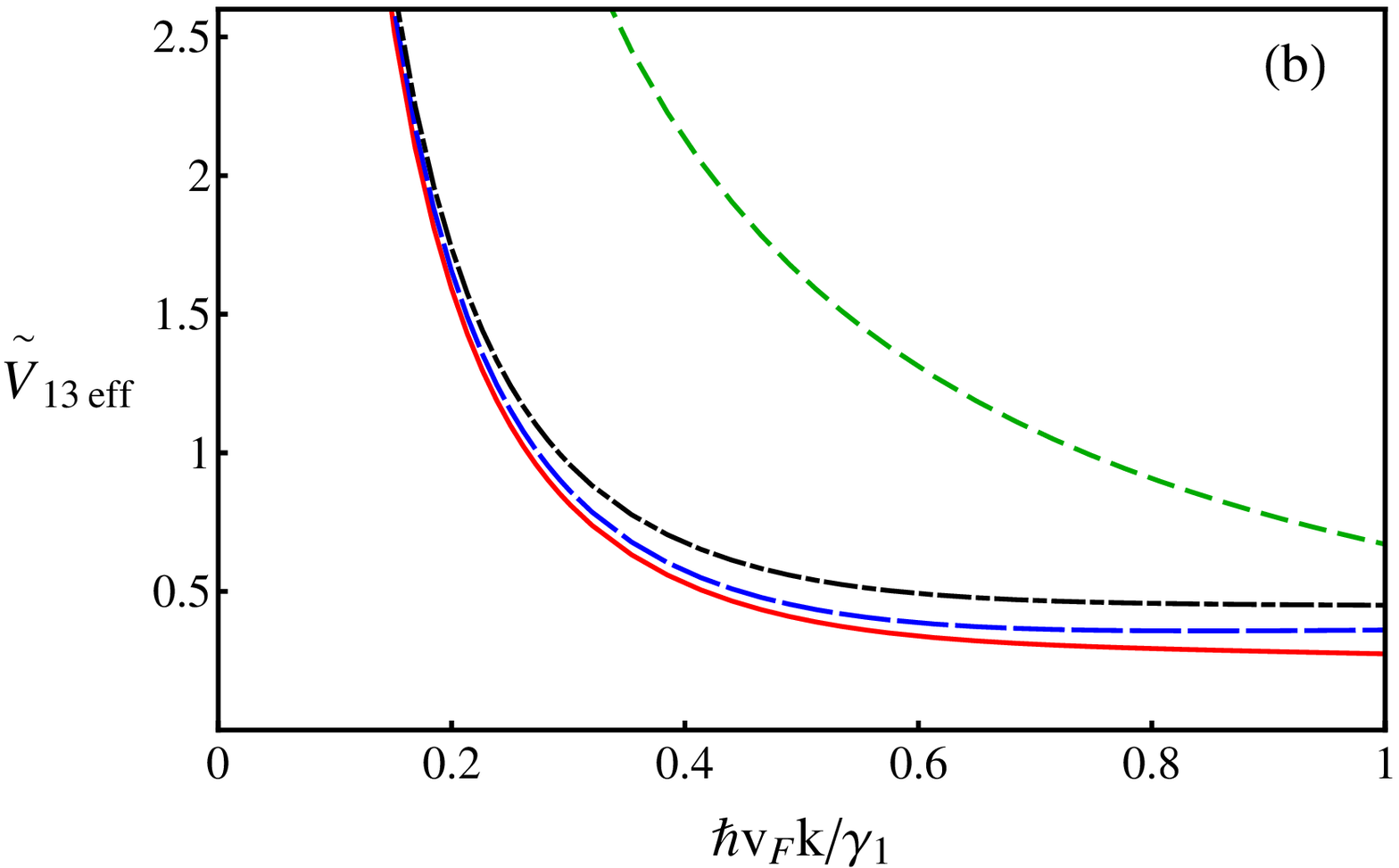}
  \caption{(Color online) The dimensionless effective interactions in gapless ABC-stacked trilayer graphene as functions of dimensionless momentum $y$ at energy $\omega=0.05\gamma_1$ for $T=0.002\gamma_{1}$, $B=1$\,T, and $\mu=0.02\gamma_{1}$. (a) $V_{eff}$ given
  by Eq. (\ref{interaction-effective}) (red solid  line), Eq. (\ref{interaction-effective_simp2})
  (blue dashed  line), and Eq. (\ref{interaction-effective_simp3}) (black dash-dotted line).
  The green dash-double-dotted line corresponds to the bare intralayer Coulomb interaction. (b)  $V_{13\,eff}$ given by Eq. (\ref{interaction-ND}) (red solid
  line), Eq. (\ref{interaction-ND_simp2}) (blue dashed line), and
  Eq. (\ref{interaction-effective_simp3}) (black dash-dotted line).
  The green dash-double-dotted line corresponds to the bare interlayer Coulomb interaction.}
\label{trilayer-effective-interactions}
\end{figure}

In this section, we analyze the dynamical screening in clean rhombohedral graphene.
For $\Gamma_{N}=0$ all denominators in Eq. (\ref{S-function}) are the same. Therefore,
\begin{eqnarray}
Z^{\lambda\lambda^{\prime}}_{NN^{\prime}}(\omega,0,\mu,T)+Z^{-\lambda^{\prime},-\lambda}_{N^{\prime}N}
(\omega,0,-\mu,T)-Z^{\lambda\lambda}_{NN}(\omega,0,\mu,T)-Z^{-\lambda^{\prime},-
\lambda^{\prime}}_{N^{\prime}N^{\prime}}(\omega,0,-\mu,T)
=n_{\rm F}(\lambda^{\prime} M_{N^{\prime}})-n_{\rm F}(\lambda M_{N}).
\end{eqnarray}
By using this result, we find that Eqs. (\ref{P_11-frequency}) and (\ref{P_1n-frequency}) imply
the following dynamical polarization functions in clean rhombohedral graphene:
\begin{equation}
\Pi_{11}(\omega,\mathbf{k})=-\frac{e^{-y}}{4\pi l^{2}} \sum\limits_{N,N^{\prime}=0}^{\infty}(-1)^{N+N^{\prime}} \sum\limits_{\lambda,
\lambda^{\prime}=\pm 1}\frac{n_{\rm F}(\lambda M_{N})-n_{\rm F}(\lambda^{\prime}M_{N^{\prime}})}{M_{N}M_{N^{\prime}}
(\lambda M_{N}-\lambda^{\prime}M_{N^{\prime}}-\omega)}\nonumber
\end{equation}
\begin{equation}
\times\left[(M_{N}-\lambda\Delta)(M_{N^{\prime}}-\lambda^{\prime}\Delta)L^{N-N^{\prime}}_{N^{\prime}}
(y)L^{N^{\prime}-N}_{N}(y)+(M_{N}+\lambda\Delta)(M_{N^{\prime}}+\lambda^{\prime}\Delta)
L^{N-N^{\prime}}_{N^{\prime}-n}(y)L^{N^{\prime}-N}_{N-n}(y)\right],
\label{clean-P_11}
\end{equation}
\begin{eqnarray}
\label{clean-P_1n}
\Pi_{1n}(\omega,\mathbf{k})&=&-\frac{e^{-y}\mathcal{E}^{2}_{n}}{2\pi l^{2}} \sum\limits_{N,N^{\prime}=0}^{\infty}(-1)^{N+N^{\prime}}
\frac{N!}{(N-n)!}L^{N^{\prime}-N}_{N}(y)L^{N-N^{\prime}}_{N^{\prime}-n}(y)\nonumber\\
&\times&\sum\limits_{\lambda,\lambda^{\prime}=\pm 1} \frac{\lambda\lambda^{\prime}
(n_{\rm F}(\lambda M_{N})-n_{\rm F}(\lambda^{\prime}M_{N^{\prime}}))}{M_{N}M_{N^{\prime}}
(\lambda M_{N}-\lambda^{\prime}M_{N^{\prime}}-\omega)}.
\end{eqnarray}
\begin{figure}[ht]
  \centering
  \includegraphics[scale=0.45]{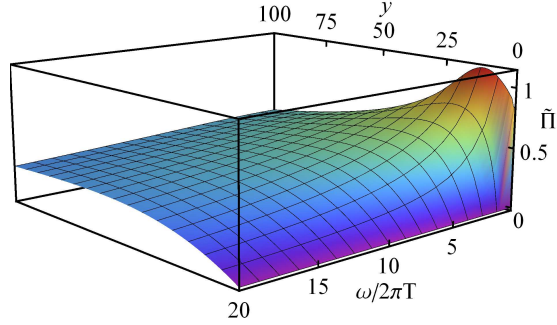}
  \caption{(Color online) The dimensionless dynamical polarization function $\tilde{\Pi}(\omega,y)$
  for the QAH state gap $\Delta=0.01\gamma_{1}$ in clean, $\Gamma_N=0$, ABC-stacked trilayer
  graphene in the magnetic field $B=1$\,T for $\mu=2\Delta$ and $T=0.5\Delta$.}
\label{trilayer-dynamical-polarization}
\end{figure}
The three-dimensional ($3D$) plot of dimensionless dynamical polarization function $\tilde{\Pi}=\frac{2\hbar^2
v_{\rm F}^2}{\gamma_1}\Pi$ for ABC-stacked trilayer graphene is presented in Fig. \ref{trilayer-dynamical-polarization}. As is seen, the polarization is maximal at small
 frequency and momentum and decreases with the increase of $\omega$.
In Fig. \ref{trilayer-effective-interactions} we
plot the dimensionless screened Coulomb interactions $\tilde{V}_{eff}(\omega,k)={V}_{eff}(\omega,k)
\kappa \gamma_{1}/(2\pi e^2 \hbar v_{\rm F})$ and $\tilde{V}_{13\,eff}(\omega,k)=
{V}_{13\,eff}(\omega,k)\kappa \gamma_{1}/(2\pi e^2 \hbar v_{\rm F})$ in gapless
trilayer graphene at fixed energy $\omega=0.05\gamma_1$ with $\kappa=5$.
For comparison, we display also the dimensionless bare intralayer, $\tilde{V}(k)={V}(k)\kappa \gamma_{1}/(2\pi e^2 \hbar v_{\rm F})$,  and interlayer, $\tilde{V}_{13}(k)={V}_{13}(k)\kappa \gamma_{1}/(2\pi e^2 \hbar v_{\rm F})$, Coulomb interactions.
As one can see, the dynamical polarization strongly reduces the effective potentials.

Simplified analytical expressions can be obtained for small momenta and in the strong-magnetic-field limit. At small momenta, the general expression for the polarization function $\Pi(\omega,y)
=\Pi_{11}(\omega,y)+\Pi_{1n}(\omega,y)$ simplifies to
\begin{equation}
\Pi(\omega,y)\simeq A+By,
\end{equation}
where
\begin{eqnarray}
A&=&-\frac{2i}{\pi l^2\omega}\sum\limits_{N=0}^\infty\frac{\Gamma_N}{\omega+2i\Gamma_N}
{\sum_{\lambda=\pm1}}^{\prime}\left[Z^{\lambda\lambda}_{NN}(\omega,\Gamma,\mu,T)+(\mu\to-\mu)\right],\\
B&=&\frac{1}{\pi l^{2}}\sum\limits_{N=0}^{\infty}{\sum\limits_{\lambda=\pm 1}}^{\prime}
\left[2N+1-n\left(1+\lambda\frac{\Delta}{M_{N}}\right)\right]\mathcal{S}(N,N,\lambda,\lambda,\omega)
\label{coeff-A}
\nonumber\\
&-&\frac{1}{4\pi l^{2}}\sum\limits_{N=0}^{\infty}{\sum\limits_{\lambda,\lambda^{\prime}=\pm 1}}^{
\prime \prime}\left(\mathcal{S}(N,N+1,\lambda,\lambda^{\prime},\omega)+
\mathcal{S}(N+1,N,\lambda^{\prime},\lambda,\omega)\right)\nonumber\\
&\times&\left[2(N+1)\left(1+\lambda\lambda^{\prime}\frac{M_{N}}{M_{N+1}}\right)
-n\left(1+\lambda\frac{\Delta}{M_{N}}\right)\left(1+\lambda^{\prime}\frac{\Delta}{M_{N+1}}
\right)\right],
\label{coeff-B}
\end{eqnarray}
the notations $\sum',\sum''$ are explained after Eq. (\ref{coeff-b}) and the functions
$Z$ and ${\cal S}$ are given in Appendix \ref{summation}. At zero scattering rate and finite
$\omega$ the coefficient $A$ vanishes, while the coefficient $B$ takes the form
\begin{eqnarray}
B&=&-\frac{1}{2\pi l^{2}}\sum\limits_{N=0}^{\infty}{\sum\limits_{\lambda,\lambda^{\prime}
=\pm 1}}^{\prime \prime}\frac{(\lambda M_{N}-\lambda^{\prime}M_{N+1})(n_{\rm F}(\lambda M_{N})
-n_{\rm F}(\lambda^{\prime}M_{N+1}))}{(\lambda M_{N}-\lambda^{\prime}M_{N+1})^2-\omega^2}
\nonumber\\
&\times&\left[2(N+1)\left(1+\lambda\lambda^{\prime}\frac{M_{N}}{M_{N+1}}\right)-
n\left(1+\lambda\frac{\Delta}{M_{N}}\right)\left(1+\lambda^{\prime}\frac{\Delta}{M_{N+1}}\right)
\right].
\end{eqnarray}
On the other hand, the limit $\omega\to0$ of the coefficient $A$ at finite $\Gamma_N$ is given
by Eq. (\ref{Pi(0,0)}). One can see that the static long-wavelength polarization function
$\Pi(0,0)$ in the magnetic field does not depend on the order of taking limits $\omega\to 0$ and $\mathbf{k}\to0$, unlike in the absence of the magnetic field.

The strong-magnetic-field limit $(l\to0)$ of the polarization function $\Pi(\omega,\mathbf{k})$
depends on the ratio between the scattering rate and frequency. For $\Gamma_N/\omega\neq0$,
the main contribution comes from the first $n$ Landau levels and is given by the expression
\begin{equation}
\Pi_{l\rightarrow 0}(\omega,y)=-\frac{1}{\pi^2 l^2}\sum\limits_{N=0}^{n-1}\frac{\Gamma_{N}}{\omega(\omega+2i\Gamma_{N})}
\sum\limits_{\lambda=\pm 1}\left[\psi\left(\frac{1}{2}+\frac{\lambda(\mu+\Delta)+\omega+i\Gamma_{N}}{2\pi i T}\right)
-\psi\left(\frac{1}{2}+\frac{\lambda(\mu+\Delta)+i\Gamma_{N}}{2\pi i T}\right)\right].
\end{equation}
This contribution vanishes in clean rhombohedral graphene for nonzero $\omega$.  In this case,
the leading at $\mathbf{k}\to 0$ term has the form
\begin{equation}
\Pi_{l\rightarrow 0}(\omega,y)=\frac{\mathbf{k}^2n^2}{2\pi \mathcal{E}_{n}}
\sum\limits_{N=n-1}^{\infty}\sqrt{\frac{(N-n+1)!}{N!}}\frac{1}{\left(\sqrt{N+1}+\sqrt{N-n+1}
\right)^{3}},
\end{equation}
which is equivalent to the static long-wavelength limit of the polarization function in clean
gapless graphene at zero temperature in the case when only $n$ lower Landau levels are
filled.

\section{Summary}
\label{summary}

In this paper, neglecting the trigonal warping effects and using the low-energy model, we
have derived analytical expressions for the one-loop
dynamical polarization functions $\Pi_{11}(\omega,\mathbf{k})$ and $\Pi_{1n}(\omega,\mathbf{k})$
in ABC-stacked $n$-layer graphene at finite temperature, chemical potential, quasiparticle gap,
in the presence of a magnetic field and taking into account the finite width of Landau levels.
The general results are given in terms of the digamma functions and generalized Laguerre
polynomials and have the form of double sums over Landau levels given by Eqs. (\ref{P_11-frequency}) and (\ref{P_1n-frequency}). These equations are a generalization
of the corresponding expressions for the polarization function in monolayer graphene obtained in Ref. \cite{Gusynin-Pyatkovskiy}. We  analyzed the intralayer and interlayer screened Coulomb
potentials in chiral multilayer graphene and found that, for a number of layers less than
or equal to eight, they are very well approximated by the simple expression where only the dynamical polarization function $\Pi=\Pi_{11}+\Pi_{1n}$ is present. The derived  polarization functions
can be used to calculate the dispersion relation and the decay rate of magnetoplasmons as functions
of temperature, impurity rate, and magnetic field.

We found that the magnetic field qualitatively changes the behavior of the polarization functions
at small wave vectors $|\mathbf{k}| \le l^{-1}$. Indeed, while the static polarization function $\Pi(0,\mathbf{k})$ diverges as $\mathbf{k} \to 0$ for $n \ge 3$ in the gapless case without
magnetic field \cite{Min-polarization}, it is no longer divergent in the presence of a magnetic
field [see, Eq. (\ref{staticPi-k-to0})].

The long-range behavior of the screened static Coulomb potential in a magnetic field is governed
by the Thomas-Fermi wave vector $k_F=(4\pi e^2/\kappa)\Pi(0,0)$, and the strength of screening
is found to oscillate as a function of chemical potential or magnetic field. If both scattering
rate and temperature are small, these oscillations turn into a sequence of delta-like functions
and, for integer fillings, the screening is absent. This suggests a possibility to tune, for example,
the strength of the impurity  by controlling the occupation of Landau level states with a gate voltage.
Such a possibility was demonstrated  recently in experiments in monolayer graphene \cite{Andrei}
and it would be interesting to observe the corresponding behavior in multilayer graphene.

\begin{acknowledgments}We are grateful to P.~Pyatkovskiy for useful comments and suggestions.
The authors acknowledge the support of the European IRSES Grant SIMTECH No. 246937.
 The work of E.V.G. and V.P.G. was also supported by the
Science and Technology Center in Ukraine (STCU) and the National Academy of Sciences of Ukraine
(NASU) within the framework of the Targeted Research \& Development Initiatives (TRDI) Program
under Grant No. 5716-2  "Development of Graphene Technologies and Investigation of Graphene-based
Nanostructures for Nanoelectronics and Optoelectronics".
\end{acknowledgments}

\appendix\section{Fermion propagator}
\label{3}

We determine the Green's function through the expansion over the eigenfunctions
($\Delta_{\xi s}\equiv \Delta$),
\begin{eqnarray}
&&S(\mathbf{r},\mathbf{r}^{\prime};\omega)=\sum\limits_{N=0,\alpha=\pm 1}^{\infty}
\int_{-\infty}^{\infty} dk\frac{\Psi(N,k,\alpha;x,y)\otimes\Psi^{\dagger}(N,k,\alpha;x^{\prime},y^{\prime})}
{\omega-E_{N\alpha}}\nonumber\\
&&=\frac{1-\tau^3}{2(\omega+\Delta)}
\int_{-\infty}^{\infty}d k\,e^{i k(y-y^{\prime})}[u_{0}(\eta)u_{0}(\eta^{\prime})+\ldots
+u_{n-1}(\eta)u_{n-1}(\eta^{\prime})]+\sum\limits_{N=n}^{\infty}
\frac{1}{\omega^{2}-M^{2}_{N}}\int_{-\infty}^{\infty}d k\,e^{i k(y-y^{\prime})}\nonumber\\
&&\times\left[(\omega+\tau^3 \Delta)\left(\begin{array}{cc}u_{N-n}(\eta)u_{N-n}
(\eta^{\prime})&0\\0&u_{N}(\eta)u_{N}(\eta^{\prime})
\end{array}\right)-\xi^n \sqrt{M^{2}_{N}-\Delta^{2}}(-i \tau^3)^n \left(\begin{array}{cc}0
&u_{N-n}(\eta)u_{N}(\eta^{\prime})\\
u_{N}(\eta)u_{N-n}(\eta^{\prime})&0
\end{array}\right)\right].\nonumber\\
\end{eqnarray}
Using the following formula 7.378 in Ref. \cite{GR},
\begin{equation}
\int\limits_{-\infty}^\infty
e^{-x^2}H_m(x+y)H_n(x+z)dx=2^n\pi^{1/2}m!z^{n-m}L_m^{n-m}(-2yz),\quad m\le n,
\end{equation}
where $L_{n}^{\nu}(z)$ are the generalized Laguerre polynomials, we can calculate the integral over $k$,
\begin{eqnarray}
\int\limits_{-\infty}^{\infty}d k\,e^{ik(y-y^{\prime})}u_{m}(\eta)u_{n}(\eta^{\prime}) &=&
\frac{1}{2\pi l^{2}}\exp\left(-\frac{(\mathbf{r}-
\mathbf{r}^{\prime})^{2}}{4l^{2}}-i\frac{(x+x^{\prime})(y-y^{\prime})}{2l^{2}}\right)\nonumber\\
&\times&\sqrt{\frac{2^{n}m!}{2^{m}n!}}\left(-\frac{(x-x^{\prime})-i(y- y^{\prime})}{2l}
\right)^{n-m}L^{n-m}_{m}\left(\frac{(\mathbf{r}-\mathbf{r}^{\prime})^{2}}{2l^{2}}\right).
\end{eqnarray}
The Green's function can be represented as follows:
\begin{equation}
S(\mathbf{r},\mathbf{r}^{\prime};\omega)=\exp\left(i\Phi\right)\tilde{S}(\mathbf{r}
-\mathbf{r}^{\prime};\omega),
\end{equation}
where $\tilde{S}(\mathbf{r}-\mathbf{r}^{\prime};\omega)$ is the translation invariant part and
$\Phi=\frac{ie}{\hbar c}\int \limits_{\bf r}^{{\bf r}^\prime}A_i^{ext}(z)dz_{i}$ is the Schwinger phase, which is not translation invariant. We find
\begin{eqnarray}
\label{translation-invariant-part}
\tilde{S}(\mathbf{r}-\mathbf{r}^{\prime};\omega)&=&\frac{1}{2\pi l^{2}}\,e^{-z/2}
\left\{\sum\limits_{N=0}^{\infty}\frac{1}{\omega^{2}-M^{2}_{N}}\left[(\omega+\tau^3\Delta)
\left(\frac{1-\tau^3}{2}L_{N}\left(z\right)+\frac{1+\tau^3}{2}L_{N-n}\left(z\right)\right)\right.
\right.\nonumber\\
&-&\left.\left.\frac{i^n \xi^n a_n }{l^{2n}}L^{n}_{N-n}(z)\left(\begin{array}{cc}0
& [(x-x^{\prime})-i(y-y^{\prime})]^{n} \\
\left[(x-x^{\prime})+i(y-y^{\prime})\right]^{n} &0\end{array}\right)\right]\right\}, \quad
z=\frac{(\mathbf{r}-\mathbf{r}^{\prime})^{2}}{2l^{2}}.
\end{eqnarray}

For completeness, we present here the momentum space expression for the translation
invariant part of the fermion Green's function
\begin{equation}
\tilde{S}(\mathbf{k},\omega)=2e^{-\mathbf{k}^{2}l^{2}}\left\{
\sum\limits_{N=0}^{\infty}\frac{(-1)^{N}}{\omega^{2}-M^{2}_{N}}
\left[(\omega+\tau^3\Delta)\left(\frac{1-\tau^3}{2}L_{N}(2\mathbf{k}^{2}l^{2})
+(-1)^n \frac{1+\tau^3}{2}L_{N-n}(2\mathbf{k}^{2}l^{2})\right)\right.\right.\nonumber
\end{equation}
\begin{equation}
-\left.\left.(-1)^n (2\xi)^n a_n L^{n}_{N-n}(2\mathbf{k}^{2}l^{2})\left(\begin{array}{cc}0
&(k_{x}-ik_{y})^{n} \\
(k_{x}+ik_{y})^{n} &0 \end{array}\right)\right]\right\}\nonumber
\end{equation}
\begin{equation}
=2e^{-\mathbf{k}^{2}l^{2}}\sum\limits_{N=0}^{\infty}\frac{(-1)^{N}}{\omega^{2}-M^{2}_{N}}
\left(\begin{array}{cc}(\omega+\Delta)(-1)^n L_{N-n}(2\mathbf{k}^{2}l^{2})
&-(-1)^n (2\xi)^n a_n (k_{x}-ik_{y})^{n} L^{n}_{N-n}(2\mathbf{k}^{2}l^{2}) \\
-(-1)^n (2\xi)^n a_n (k_{x}+ik_{y})^{n} L^{n}_{N-n}(2\mathbf{k}^{2}l^{2}) &
(\omega-\Delta)L_{N}(2\mathbf{k}^{2}l^{2})\end{array}\right)
\label{FT-propagator}
\end{equation}
(by definition $L_{-m}\equiv 0,\ m>0$).

\section{Summation over Matsubara frequencies}
\label{summation}

In order to find the polarization functions in momentum space, we use the following integrals:
\begin{eqnarray}
I^{(n)}_{N,K}(y)&=&\int d^{2}\mathbf{r}\,e^{-i\mathbf{k}\mathbf{r}}e^{-z}L^{n}_{N-n}(z)L^{n}_{K-n}(z)
r^{2n}\nonumber\\
&=&4\pi l^{2} (2l^{2})^{n} (-1)^{n} \frac{N!}{(N-n)!}\int\limits_{0}^{\infty}dx\,x\,e^{-x^{2}}\,
J_{0}\left(\sqrt{2{k^{2}l^{2}}}x\right)L^{-n}_{N}(x^{2})L^n_{K-n}(x^{2})
\nonumber\\
&=&2\pi l^{2}(-1)^{N+K}(2l^{2})^{n} \frac{N!}{(N-n)!} e^{-y}L^{K-N}_{N}
\left(y\right)L^{N-K}_{K-n}\left(y\right),\quad z=\frac{\mathbf{r}^2}{2l^2},\quad y=\frac{\mathbf{k}^{2}l^{2}}{2},
\label{second-integral}
\end{eqnarray}
where we used Eq. (7.422.2) in Ref. \cite{GR} and the following property of the Laguerre
polynomials for $l,\, k \geq 0$:
\begin{equation}
\label{property-Laguerre}
L^{k}_{l}(x)=(-x)^{-k}\frac{(l+k)!}{l!}L^{-k}_{l+k}(x)\,\,\Longrightarrow\,\,x^{n}L^{n}_{N-n}(x)
=(-1)^{n}\frac{N!}{(N-n)!}L^{-n}_{N}(x),
\end{equation}
where $l,\, k\geq 0$. (For more details of calculations of similar integrals, see Appendix A in Ref. \cite{Gusynin-Pyatkovskiy}.)

The dynamical polarization functions (\ref{polarization-imp-1n}) at finite temperature, chemical potential, and nonzero width $\Gamma_N$ take
the form
\begin{eqnarray}
\Pi_{11}(i\Omega_{p}, \mathbf{k})&=&-\frac{T}{\pi l^{2}} \sum\limits_{m=-\infty}^{+\infty}
\sum\limits_{N,N^{\prime}=0}^{\infty}\frac{(-1)^{N+N^{\prime}}e^{-y}}
{((i\omega_{m}+\mu+i\Gamma_{N}{\rm sgn}\omega_{m})^{2}-
M^{2}_{N})((i\omega_{m-p}+\mu+i\Gamma_{N^{\prime}}{\rm sgn}
\omega_{m-p})^{2}-M^{2}_{N^{\prime}})}\nonumber\\
&\times&\left[(i\omega_{m}+\mu+i\Gamma_{N}{\rm sgn}
\omega_{m}-\Delta)(i\omega_{m-p}+\mu+i\Gamma_{N^{\prime}}{\rm sgn}\omega_{m-p}-\Delta)L^{N-
N^{\prime}}_{N^{\prime}}(y)L^{N^{\prime}-N}_{N}(y)\right.\nonumber\\
&+&\left.(i\omega_{m}+\mu+i\Gamma_{N}{\rm sgn}\omega_{m}+\Delta)(i\omega_{m-p}+\mu+i\Gamma_{N^{\prime}}{\rm sgn}\omega_{m-p}+\Delta)L^{N-
N^{\prime}}_{N^{\prime}-n}(y)L^{N^{\prime}-N}_{N-n}(y)\right],
\label{polarization-imp-matsubara-width-11}
\end{eqnarray}
\begin{eqnarray}
\label{polarization-imp-matsubara-width-1n}
\Pi_{1n}(i\Omega_{p}, \mathbf{k})&=&-\frac{T}{\pi l^{2}} \sum\limits_{m=-\infty}^{+\infty}
\sum\limits_{N,N^{\prime}=0}^{\infty}\frac{(-1)^{N+N^{\prime}}e^{-y}}
{((i\omega_{m}+\mu+i\Gamma_{N}{\rm sgn}\omega_{m})^{2}-
M^{2}_{N})((i\omega_{m-p}+\mu+i\Gamma_{N^{\prime}}{\rm sgn}\omega_{m-p})^{2}-M^{2}_{N^{\prime}})}\nonumber\\
&\times&2\left[\mathcal{E}^{2}_{n}N(N-1)\ldots(N-n+1)\right]L^{N^{\prime}-N}_{N}(y)
L^{N-N^{\prime}}_{N^{\prime}-n}(y).
\end{eqnarray}

The summation over Matsubara frequencies is performed as usual $T\sum\limits_{m=-\infty}^{+\infty}f(i\omega_{m}+\mu)
=\frac{1}{2\pi i}\oint\limits_{\gamma}n_{\rm F}(\omega)f(\omega)d\omega$, where $n_{\rm F}(x)=\left\{\exp(\frac{x-\mu}{T})+1\right\}^{-1}$ is
the Fermi distribution, with subsequent deformation of the contour over the poles of the function $f(\omega)$.
It is convenient also to use the identities
\begin{eqnarray}
\frac{1}{x^2-b^2}&=&\frac{1}{2b}\left(\frac{1}{x-b}-\frac{1}{x+b}\right)=\sum\limits_{\lambda=\pm 1}\frac{\lambda}{2b(x-\lambda b)},\nonumber\\
\frac{x+a}{x^2-b^2}&=&\frac{b+a}{2b}\frac{1}{x-b}+\frac{b-a}{2b}\frac{1}{x+b}=\sum\limits_{\lambda=\pm 1}\frac{b+\lambda a}{2b(x-\lambda b)}.
\end{eqnarray}
Then we have
\begin{eqnarray}
S_1&=&T\sum\limits_{m=-\infty}^{\infty}\frac{1}{((i\omega_{m}+\mu+i\Gamma_{N}{\rm sgn}
\omega_{m})^{2}-M^{2}_{N})((i\omega_{m-p}+\mu+i\Gamma_{N^{\prime}}{\rm sgn}
\omega_{m-p})^{2}-M^{2}_{N^{\prime}})}\nonumber\\
&=&\sum\limits_{\lambda,\lambda^{\prime}=\pm1}
\frac{\lambda\lambda^{\prime}\,\mathcal{S}}{4M_{N}M_{N^{\prime}}},
\end{eqnarray}
\begin{eqnarray}
S_2&=&T\sum\limits_{m=-\infty}^{\infty}\frac{(i\omega_{m}+\mu+i\Gamma_{N}{\rm sgn}
\omega_{m}+\Delta)(i\omega_{m-p}+\mu+i\Gamma_{N^{\prime}}{\rm sgn}
\omega_{m-p}+\Delta)}{((i\omega_{m}+\mu+i\Gamma_{N}{\rm sgn}
\omega_{m})^{2}-M^{2}_{N})((i\omega_{m-p}+\mu+i\Gamma_{N^{\prime}}{\rm sgn}
\omega_{m-p})^{2}-M^{2}_{N^{\prime}})}\nonumber\\
&=&\sum\limits_{\lambda,\lambda^{\prime}=\pm 1}\frac{(M_{N}+\lambda\Delta)
(M_{N^{\prime}}+\lambda^{\prime}\Delta)\,\mathcal{S}}{4M_{N}M_{N^{\prime}}},
\end{eqnarray}
where
\begin{equation}
\mathcal{S}=T\sum\limits_{m=-\infty}^{\infty}\frac{1}{(i\omega_{m}+\mu+i\Gamma_{N}
{\rm sgn}\omega_{m}-\lambda M_{N})(i\omega_{m-p}+\mu+i\Gamma_{N^{\prime}}{\rm sgn}\omega_{m-p}-\lambda^{\prime} M_{N^{\prime}})}.
\end{equation}
To evaluate the sum we expand it in terms of partial fractions and, using the summation formula,
\begin{equation*}
\sum\limits_{m=0}^{\infty}\left(\frac{1}{m+a}-\frac{1}{m+b}\right)=\psi(b)-\psi(a),
\end{equation*}
we obtain
\begin{eqnarray}
\mathcal{S}(N,N',\lambda,\lambda',i\Omega_p)&=&
-\frac{Z^{\lambda\lambda^{\prime}}_{NN^{\prime}}(i\Omega_{p},\Gamma,\mu,T)}{\lambda M_{N}-\lambda^{\prime}M_{N^{\prime}}-i\Omega_{p}-
i\left(\Gamma_{N}-\Gamma_{N^{\prime}}\right)}
-\frac{Z^{-\lambda^{\prime},-\lambda}_{N^{\prime}N}(i\Omega_{p},\Gamma,-\mu,T)}{\lambda
M_{N}-\lambda^{\prime}M_{N^{\prime}}-i\Omega_{p}+i\left(\Gamma_{N}-\Gamma_{N^{\prime}}\right)}
\nonumber\\
&+&\frac{Z^{\lambda\lambda}_{NN}(i\Omega_{p},\Gamma,\mu,T)
+Z^{-\lambda^{\prime},-\lambda^{\prime}}_{N^{\prime}N^{\prime}}(i\Omega_{p},\Gamma,-\mu,T)}
{\lambda M_{N}-\lambda^{\prime}M_{N^{\prime}}-i\Omega_{p}-i\left(\Gamma_{N}+\Gamma_{N^{\prime}}
\right)},
\label{S-function}
\end{eqnarray}
where we introduced a new function
\begin{equation}
Z^{\lambda\lambda^{\prime}}_{NN^{\prime}}(\omega,\Gamma,\mu,T)=\frac{1}{2\pi i}
\left(\psi\left(\frac{1}{2}+\frac{\mu+\omega+i\Gamma_{N}-\lambda
M_{N}}{2\pi i T}\right)-\psi\left(\frac{1}{2}+\frac{\mu+i\Gamma_{N^{\prime}}-\lambda^{\prime} M_{N^{\prime}}}{2\pi i T}\right)\right)
\label{Z-definition}
\end{equation}
in terms of the digamma function $\psi(z)$.
When deriving Eq. (\ref{S-function}) we used also the following relation
\begin{equation}
\label{Z}
Z^{\lambda^{\prime}\lambda^{\prime}}_{N^{\prime}N^{\prime}}(-i\Omega_p,-\Gamma,
\mu,T)=Z^{-\lambda^{\prime},-\lambda^{\prime}}_{N^{\prime}N^{\prime}}(i\Omega_p,\Gamma,-\mu,T),
\end{equation}
which is easily proved by making use of the formula
\begin{equation}
\label{property_digamma}
\psi(1-z)=\psi(z)+\pi \cot(\pi z).
\end{equation}
Thus, we arrive at the equations
\begin{equation}
\Pi_{11}(i\Omega_{p},y)=-\frac{e^{-y}}{4\pi l^{2}} \sum\limits_{N,N^{\prime}=0}^{\infty}\frac{(-1)^{N+N^{\prime}}}{M_{N}M_{N^{\prime}}}
\sum\limits_{\lambda,\lambda^{\prime}=\pm 1}\mathcal{S}(N,N',\lambda,\lambda',i\Omega_p)\nonumber\\
\end{equation}
\begin{equation}
\times\left[(M_{N}-\lambda\Delta)(M_{N^{\prime}}-\lambda^{\prime}\Delta)L^{N-N^{\prime}}_{N^{\prime}}
(y)L^{N^{\prime}-N}_{N}(y)+(M_{N}+\lambda\Delta)(M_{N^{\prime}}+\lambda^{\prime}\Delta)
L^{N-N^{\prime}}_{N^{\prime}-n}(y)L^{N^{\prime}-N}_{N-n}(y)\right],
\label{most_general_P_11}
\end{equation}
\begin{equation}
\label{most_general_P_1n}
\Pi_{1n}(i\Omega_{p},y)=-\frac{e^{-y}\mathcal{E}^{2}_{n}}{2\pi l^{2}}\sum\limits_{N,N^{\prime}=0}^{\infty}
\frac{(-1)^{N+N^{\prime}}}{M_{N}M_{N^{\prime}}}\frac{N!}{(N-n)!}L^{N^{\prime}-N}_{N}
(y)L^{N-N^{\prime}}_{N^{\prime}-n}(y)\sum\limits_{\lambda,\lambda^{\prime}=\pm 1} \lambda\lambda^{\prime}\mathcal{S}(N,N',\lambda,
\lambda',i\Omega_p).
\end{equation}
Making the analytical continuation from Matsubara frequencies by replacing $i\Omega_p
\rightarrow\omega+i0$, we obtain Eqs. (\ref{P_11-frequency}) and (\ref{P_1n-frequency})
in the main text.

\section{The polarization functions in the static limit}
\label{static-lim}

Taking the limit $\omega\rightarrow 0$ in Eqs. (\ref{P_11-frequency}) and (\ref{P_1n-frequency})
and using Eq. (\ref{Z-definition}), it is not difficult to see that the third term in Eq. (\ref{S-function}) vanishes. Then, for $\lambda N\neq \lambda^{\prime} N^{\prime}$,
$\mathcal{S}$ is
\begin{eqnarray}
\mathcal{S}(\lambda N\neq \lambda^{\prime} N^{\prime})=-\frac{1}{\pi}\Im m\left\{\frac{\psi\left(\frac{1}{2}+\frac{\Gamma_{N}}{2\pi
T}+\frac{\mu-\lambda M_{N}}{2\pi i T}\right)-\psi\left(\frac{1}{2}+\frac{\Gamma_{N^{\prime}}}{2\pi T}+\frac{\mu-\lambda^{\prime} M_{N^{\prime}}}
{2\pi i T}\right)}{\lambda M_{N}-\lambda^{\prime}M_{N^{\prime}}-i\left(\Gamma_{N}-\Gamma_{N^{\prime}}\right)}\right\}
\end{eqnarray}
For $\lambda N=\lambda^{\prime} N^{\prime}$, Eq. (\ref{S-function}) implies for $\omega\to0$
\begin{equation}
{\cal S}=-\frac{1}{2\pi^2 T}\Re e\,\psi^{\prime}\left(\frac{1}{2}+\frac{\Gamma_{N}}{2\pi T}+\frac{\mu-\lambda M_{N}}{2\pi i T}\right).
\end{equation}
Thus, we have the following static polarization functions:
\begin{eqnarray}
\label{static_P_11}
\Pi_{11}(0,y)&=&\frac{e^{-y}}{8\pi^3 l^{2}T} \sum\limits_{N=0}^{\infty}\sum\limits_{\lambda=\pm 1}\frac{1}{M_{N}^2}\Re
e\,\psi^{\prime}\left(\frac{1}{2}+\frac{\Gamma_{N}}{2\pi T}+\frac{\mu-\lambda M_{N}}{2\pi i T}\right)\left[(M_{N}-\lambda\Delta)^2\left(L_{N}
(y)\right)^2+(M_{N}+\lambda\Delta)^2\left(L_{N-n}(y)\right)^2\right]\nonumber\\
&+&\frac{e^{-y}}{4\pi^2 l^{2}} \underset{\lambda N\neq \lambda^{\prime} N^{\prime}}{ \sum\limits_{N,N^{\prime}=0}^{\infty} \sum\limits_{\lambda,
\lambda^{\prime}=\pm 1}}\frac{y^{|N-N^{\prime}|}}{M_{N}M_{N^{\prime}}}\Im m\left\{\frac{\psi\left(\frac{1}{2}+\frac{\Gamma_{N}}{2\pi T}+\frac{\mu-\lambda M_{N}}
{2\pi i T}\right)-\psi\left(\frac{1}{2}+\frac{\Gamma_{N^{\prime}}}{2\pi T}+\frac{\mu-\lambda^{\prime} M_{N^{\prime}}}{2\pi i T}\right)}{\lambda
M_{N}-\lambda^{\prime}M_{N^{\prime}} -i\left(\Gamma_{N}-\Gamma_{N^{\prime}}\right)}\right\}
\nonumber\\
&\times&\left[(M_{N}-\lambda\Delta)(M_{N^{\prime}}-\lambda^{\prime}\Delta)\frac{(N_{<})!}{(N_{>})!}
\left(L^{|N-N^{\prime}|}_{N_{<}}(y)\right)^2\hspace{-1mm}+\hspace{-1mm}(M_{N}+\lambda\Delta)
(M_{N^{\prime}}+\lambda^{\prime}
\Delta)\frac{(N_{<}-n)!}{(N_{>}-n)!}\left(L^{|N-N^{\prime}|}_{N_{<}-n}(y)\right)^2\right],\\
\Pi_{1n}(0,y)&=&\frac{e^{-y}}{8\pi^3 l^{2} T}\sum\limits_{N=n}^{\infty}\sum\limits_{\lambda=\pm 1} \frac{2\mathcal{E}^{2}_{n}}{M_{N}^2}\frac{N!}
{(N-n)!}L_{N}(y)L_{N-n}(y) \Re e\psi^{\prime}\left(\frac{1}{2}+\frac{\Gamma_{N}}{2\pi T}
+\frac{\mu-\lambda M_{N}}{2\pi i T}\right)\nonumber\\
&+&\frac{e^{-y}}{4\pi^2 l^{2}} \underset{\lambda N\neq \lambda^{\prime} N^{\prime}}{ \sum\limits_{N,N^{\prime}=n}^{\infty} \sum\limits_{\lambda,\lambda^{\prime}=\pm 1}} \frac{2\mathcal{E}^{2}_{n}\lambda\lambda^{\prime}}{M_{N}M_{N^{\prime}}}\frac{(N_{<})!}{(N_{>}-n)!}
y^{|N-N^{\prime}|}L^{|N-N^{\prime}|}_{N_{<}}(y)L^{|N-N^{\prime}|}_{N_{<}-n}(y)\nonumber\\
&\times&\Im m\left\{\frac{\psi\left(\frac{1}{2}+\frac{\Gamma_{N}}{2\pi T}+\frac{\mu-\lambda M_{N}}
{2\pi i T}\right)-\psi\left(\frac{1}{2}+\frac{\Gamma_{N^{\prime}}}{2\pi T}+\frac{\mu-
\lambda^{\prime} M_{N^{\prime}}}{2\pi i T}\right)}{\lambda
M_{N}-\lambda^{\prime}M_{N^{\prime}}-i\left(\Gamma_{N}-\Gamma_{N^{\prime}}\right)}\right\},
\label{static_P_1n}
\end{eqnarray}
where  $N_{<}=\min\{N,N^{\prime}\}$, $N_{>}=\max\{N,N^{\prime}\}$, and we used also Eq. (\ref{property-Laguerre}). We note that the last term in the second curly brackets in Eq. (\ref{static_P_11}) is considered to be equal identically zero for $N_{<}-n<0$.
The expressions (\ref{static_P_11}) and (\ref{static_P_1n}) are
obviously real functions and, by using the formulas
\begin{equation}
\Im m\,\psi\left(\frac{1}{2}+\frac{\mu-\lambda M_{N}}{2\pi i T}\right)=\frac{\pi}{2}
\left(1-2n_F\left(\lambda M_{N}\right)\right),
\end{equation}
\begin{equation}
\Re e\,\psi^{\prime}\left(\frac{1}{2}+\frac{\mu-\lambda M_{N}}{2\pi i T}\right)
=\frac{\pi^2}{2}\frac{1}{{\rm ch}^2\left(\frac{\mu-\lambda M_{N}}{2T}\right)},
\end{equation}
they give the following static polarization functions in clean rhombohedral graphene:
\begin{eqnarray}
\label{clean-static-P_11}
\Pi_{11}(0,y)&=&-\frac{e^{-y}}{4\pi l^{2}} \underset{\lambda N\neq \lambda^{\prime} N^{\prime}}
{ \sum\limits_{N,N^{\prime}=n}^{\infty}\sum\limits_{\lambda,\lambda^{\prime}=\pm 1}}\frac{1}{M_{N}M_{N^{\prime}}}\frac{n_{\rm F}(\lambda M_{N})-n_{\rm F}(\lambda^{\prime}
M_{N^{\prime}})}{\lambda M_{N}-\lambda^{\prime} M_{N^{\prime}}}\nonumber\\
&\times&y^{|N-N^{\prime}|}\left[(M_{N}-\lambda\Delta)(M_{N^{\prime}}-\lambda^{\prime}\Delta)
\frac{(N_{<})!}{(N_{>})!}\left(L^{|N-N^{\prime}|}_{N_{<}}(y)\right)^2\right.\nonumber\\
&+&\left.(M_{N}+\lambda\Delta)(M_{N^{\prime}}+\lambda^{\prime}\Delta)
\frac{(N_{<}-n)!}{(N_{>}-n)!}\left(L^{|N-N^{\prime}|}_{N_{<}-n}(y)\right)^2\right]\nonumber\\
&-&\frac{e^{-y}}{\pi l^{2}} \sum\limits_{N=n}^{\infty}\sum\limits_{\lambda=\pm 1} \left(1-\frac{\lambda\Delta}{M_{N}}\right)\frac{n_{\rm F}
(\lambda M_{N})-n_{\rm F}(-\Delta)}{\lambda M_{N}+\Delta} \sum\limits_{K=0}^{n-1}y^{N-K}\frac{K!}{N!}\left(L^{N-K}_{K}(y)\right)^2\nonumber\\
&+&\frac{e^{-y}}{16\pi l^{2} T}\sum\limits_{N=n}^{\infty}\sum\limits_{\lambda=\pm 1}
\frac{1}{{\rm ch}^2\left(\frac{\mu-\lambda M_{N}}
{2T}\right)}\left[\left(1-\frac{\lambda\Delta}{M_{N}}\right)^2(L_{N}(y))^2+\left(1+
\frac{\lambda\Delta}{M_{N}}\right)^2(L_{N-n}(y))^2\right]\nonumber\\
&+&\frac{e^{-y}}{4\pi l^{2} T}\frac{1}{{\rm ch}^2\left(\frac{\mu+\Delta}{2T}\right)}
\sum\limits_{N,N^{\prime}=0}^{n-1} y^{|N-N^{\prime}|}\frac{(N_{<})!}{(N_{>})!}
\left(L^{|N-N^{\prime}|}_{N_{<}}\right)^2,\\
\Pi_{1n}(0,y)&=&\frac{e^{-y}}{16 \pi l^{2} T}\sum\limits_{N=n}^{\infty}\sum\limits_{\lambda=\pm 1} \frac{2\mathcal{E}^{2}_{n}}{M_{N}^2}\frac{N!}
{(N-n)!}L_{N}(y)L_{N-n}(y) \frac{1}{{\rm ch}^2\left(\frac{\mu-\lambda M_{N}}{2T}\right)}\nonumber\\
&-&\frac{e^{-y}}{4\pi l^{2}} \underset{\lambda N\neq \lambda^{\prime} N^{\prime}}{ \sum\limits_{N,N^{\prime}=n}^{\infty} \sum\limits_{\lambda,
\lambda^{\prime}=\pm 1}} \frac{2\mathcal{E}^{2}_{n}\lambda\lambda^{\prime}}{M_{N}M_{N^{\prime}}}\frac{(N_{<})!}{(N_{>}-n)!}y^{|N-
N^{\prime}|}L^{|N-N^{\prime}|}_{N_{<}}(y)L^{|N-N^{\prime}|}_{N_{<}-n}(y)\nonumber\\
&\times&\frac{n_{\rm F}(\lambda M_{N})-n_{\rm F}(\lambda^{\prime} M_{N^{\prime}})}{\lambda M_{N}-\lambda^{\prime} M_{N^{\prime}}}.
\label{clean-static-P_1n}
\end{eqnarray}


\begin{thebibliography}{99}


\bibitem{Guinea} F. Guinea, A.H. Castro Neto, and N.M.R. Peres, Phys. Rev. B {\bf 73}, 245426 (2006).

\bibitem{Min} Hoghki Min and A.H. MacDonald, Phys. Rev. B {\bf 77}, 155416 (2008).

\bibitem{Barlas} Yafis Barlas, Kun Yang, and A.H. MacDonald, Nanotechnology {\bf 23}, 052001 (2012).

\bibitem{Koshino} M. Koshino and E. McCann, Phys. Rev. B {\bf 80}, 165409 (2009).

\bibitem{Sahu} F. Zhang, B. Sahu, H. Min, and A.H. MacDonald, Phys. Rev. B {\bf 82}, 035409 (2010).

\bibitem{Martin} J.~Martin, B.~E.~Feldman, R.~T.~Weitz, M.~T.~Allen, and A.~Yacoby,
Phys. Rev. Lett. {\bf 105}, 256806 (2010).

\bibitem{Weitz} R.~T.~Weitz, M.~T.~Allen, B.~E.~Feldman, J.~Martin, and A.~Yacoby,
Science {\bf 330}, 812 (2010).

\bibitem{Freitag} F.~Freitag, J.~Trbovic, M.~Weiss, and C.~Schonenberger, Phys. Rev. Lett.
{\bf 108}, 076602 (2012).

\bibitem{Velasco} J.~Velasco Jr., L. Jing, W. Bao, Y. Lee, P. Kratz, V.~Aji, M.~Bockrath,
C.N. Lau, C. Varma, R. Stillwell, D. Smirnov, F. Zhang, J.~Jung, and A.H.~MacDonald,
Nat.~Nanotechnol. {\bf 7}, 156 (2012).

\bibitem{Gillgren} Y. Lee, D. Tran, K. Myhro, J. Velasco Jr., N. Gillgren, C.N. Lau, Y. Barlas, J.M. Poumirol, D. Smirnov, and F. Guinea, arXiv:1402.6413 [cond-mat.str-el].

\bibitem{Bao} W.~Bao, L. Jing, J. Velasco Jr., Y. Lee, G. Liu, D. Tran, B. Standley, M. Aykol,
S.B. Cronin, D. Smirnov, M. Koshino, E. McCann, M. Bockrath, and C.N. Lau, Nat.~Phys.
{\bf 7}, 948 (2011).

\bibitem{Avetisyan} A.A. Avetisyan, B. Partoens, and F.M. Peeters, Phys. Rev. B {\bf 79}, 035421
(2009); {\it ibid.} Phys. Rev. B {\bf 81}, 115432 (2010).

\bibitem{Heinz} C.H. Lui, Z. Li, K.F. Mak, E. Cappelluti, and T.F. Heinz, Nat.~Phys.
{\bf 7}, 944 (2011).

\bibitem{Clapp} K. Zou, F. Zhang, C. Clapp, A.H. MacDonald, and J. Zhu, Nano~Lett. {\bf 13},
issue 2, 369 (2013).

\bibitem{McCann} M. Koshino, Phys. Rev. B {\bf 81}, 125304 (2010).

\bibitem{Russo} S. Russo, M.F. Craciun, T. Khodkov, M. Koshino, M. Yamamoto, and S. Tarucha,
In the book:  "Graphene - Synthesis, Characterization, Properties and Applications", edited by
Jian Gong (InTech, Rijeka, Croatia, 2011).

\bibitem{Rondeau}Y.~Barlas, R.~Cote, and M.~Rondeau, Phys. Rev. Lett. {\bf109}, 126804 (2012).

\bibitem{Min-polarization} Hongki Min, E.H. Hwang, and S. Das Sarma, Phys. Rev. B
{\bf 86}, 081402(R) (2012).

\bibitem{Gelderen} R. van Gelderen, R. Olsen, and C.M. Smith, Phys. Rev. B
{\bf88}, 115414 (2013).

\bibitem{Olsen} R. Olsen, R. van Gelderen, and C.M. Smith, Phys. Rev. B {\bf 87},
115414 (2013).

\bibitem{Jia} Junji Jia, E.V. Gorbar, and V.P. Gusynin, Phys. Rev. B {\bf 88}, 205428 (2013).

\bibitem{Cote} R.~C\^ot\'e, M. Rondeau, A.-M. Gagnon, and Yafis Barlas, Phys. Rev. B {\bf 86}, 125422 (2012).

\bibitem{Nakamura} M. Nakamura and L. Hirasawa, Phys. Rev. B {\bf 77}, 045429 (2008).

\bibitem{CoteBarrette} R.~C\^ot\'e and M.~Barrette, Phys. Rev. B {\bf 88}, 245445 (2013).

\bibitem{bilayer} E.V. Gorbar, V.P. Gusynin, and V.A. Miransky, Phys. Rev. B {\bf 81}, 155451
(2010).

\bibitem{Haldane} F.D.M. Haldane, {Phys. Rev. Lett.} {\bf 61}, 2015 (1988).

\bibitem{Bateman} H.~Bateman, A.~Erdelyi, {\it Higher Transcendental Functions}, V.2, (Mc Graw-Hill, New York-Toronto-London, 1953).

\bibitem{Davies} J.~H.~Davies, {\it The Physics of Low-dimensional Semiconductors: an Introduction}
(Cambridge University Press, Cambridge, 1998).

\bibitem{Andrei}A.~Luican-Mayer, M.~Kharitonov, G. Li, C.P. Lu, I.~Skachko,
Alem-Mar~B.~Gon\c{c}alves, K.~Watanabe, T.~Taniguchi, and E.Y.~Andrei, Phys. Rev. Lett. {\bf112},
036804 (2014).

\bibitem{Gusynin-Pyatkovskiy} P.K.~Pyatkovskiy and V.P.~Gusynin, Phys. Rev. B {\bf 83}, 075422 (2011).

\bibitem{FNT2014}V.P. Gusynin, V.M. Loktev, I.A. Luk'yanchuk, S.G. Sharapov, and A.A. Varlamov,
Fizika Nizkikh Temperatur {\bf40}, No. 4, 355 (2014) [Low Temp. Phys. {\bf40}, 270 (2014].

\bibitem{QAH}R.~Nandkishore and L.~Levitov, Phys. Rev. Lett. {\bf 104}, 156803 (2010).

\bibitem{LAF}H. Min, G. Borghi, M. Polini, and A.H. MacDonald, Phys. Rev. B {\bf 77}, 041407(R) (2008).

\bibitem{Kharitonov}M.~Kharitonov, Phys. Rev. B {\bf86}, 195435 (2012).

\bibitem{Roy}B.~Roy, Phys. Rev. B {\bf89}, 201401 (2014).

\bibitem{GR} I.S.~Gradsteyn and I.M.~Ryzhik, {\it Tables of Integrals, Series,
and Products}, (Academic Press, New York, 1965).






\end{thebibliography}
\end{document}